# Pareto Efficient Multi Objective Optimization for Local Tuning of Analogy Based Estimation

Mohammad Azzeh
Department of Software Engineering
Applied Science University
Amman, Jordan POBOX 166
m.y.azzeh@asu.edu.jo

Ali Bou Nassif
Department of Electrical and Computer Engineering
University of Sharjah
Sharjah, UAE
abounassif@ieee.org

Shadi Banitaan
Department of Mathematics, Computer Science and Software Engineering
University of Detroit Mercy, USA
banitash@udmercy.edu

Fadi Almasalha
Department of Computer Science
Applied Science University
Amman, Jordan POBOX 166
f_masalha@asu.edu.jo

**Abstract.**
Analogy Based Effort Estimation (ABE) is one of the prominent methods for software effort estimation. The fundamental concept of ABE is closer to the mentality of expert estimation but with an automated procedure in which the final estimate is generated by reusing similar historical projects. The main key issue when using ABE is how to adapt the effort of the retrieved nearest neighbors. The adaptation process is an essential part of ABE to generate more successful accurate estimation based on tuning the selected raw solutions, using some adaptation strategy. In this study we show that there are three interrelated decision variables that have great impact on the success of adaptation method: (1) number of nearest analogies ($k$), (2) optimum feature set needed for adaptation, and (3) adaptation weights. To find the right decision regarding these variables, one need to study all possible combinations and evaluate them individually to select the one that can improve all prediction evaluation measures. The existing evaluation measures usually behave differently, presenting sometimes opposite trends in evaluating prediction methods. This means that changing one decision variable could improve one evaluation measure while it is decreasing the others. Therefore, the main theme of this research is how to come up with best decision variables that improve adaptation strategy and thus, the overall evaluation measures without degrading the others. The impact of these decisions together has not been investigated before, therefore we propose to view the building of adaptation procedure as a multi-objective optimization problem. The Particle Swarm Optimization Algorithm (PSO) is utilized to find the optimum solutions for such decision variables based on optimizing multiple evaluation measures. We evaluated the proposed approaches over 15 datasets and using 4 evaluation measures. After extensive experimentation we found that: (1) predictive performance of ABE has noticeably been improved, (2) optimizing all decision variables together is more efficient than ignoring any one of them. (3) Optimizing decision variables for each project individually yield better accuracy than optimizing them for the whole dataset.

## 1. Introduction

One of the key challenges in software industry is how to obtain the accurate estimation of the development effort, which is particularly important for risk evaluation, resource scheduling as well as progress monitoring [3][15][28]. This importance is clearly portrayed through proposing a vast variety of estimation models in the past years [47]. Inaccuracies in estimations lead to problematic results for both software industry and customers. In one hand the underestimation results in approval of projects that will exceed their planned budgets, while on the other hand the overestimation causes waste of resources and misses opportunities to offer funds for other projects in future [31]. Software effort estimation has been extensively studied in literature since 70's but they have suffered from common problems such as very large performance deviations as well as being highly dataset dependent [15]. These models can be classified into two main categories: a) Expert Judgment, and b) Learning Oriented models. The former proposes making use of the experiences of human experts whereas, the latter usually generates estimates based on learning methods. The latter has two distinct advantages over the former such that they have capability to model complex set of relationships between dependent variable and the independent variables, and they are capable to learn from historical project data [2][27].

In the recent years, a significant research effort was put into utilizing various machine learning (ML) algorithms as a complementary or as a replacement to previous methods [14][17][25][31]. Although they generate successful results in certain datasets, ML algorithms suffered from local tuning problems when they were to be applied in another settings, i.e. they need to be tuned to local data for high accuracy values[15][33]. ML methods have an extremely large space of configuration possibilities [15][42][43]. When we consider configuration possibilities of ML methods induced on different datasets, each method has its own characteristics, so it is not a surprise to see contradictory results [6][9][30][35]. Finding the best estimation model was under a thorough investigation of many comparative studies that attempted to rank and categorize those models based on the quality of estimates they produce [25][27]. Unfortunately, there are no consensus conclusions between these studies on which technique is the best. The most factors that contradict their findings seem to be the error measures [25], datasets preprocessing and their inherent characteristics [3], and finally, the experimental methodology [15].

This paper focuses mainly on Analogy-Based Estimation method (hereafter ABE). The idea behind ABE method is rather simple such that a new project's effort can be estimated by reusing project efforts about similar, already documented projects in a dataset, where in the first step, one has to identify the projects which contain the most similar features. Since the utility of a project cannot be evaluated directly, similarity between project descriptions is used as a heuristic approach to retrieve the projects' effort [22]. We study ABE for several reasons: a) it reflects human reasoning, b) it works with spare data and complex domains, and c) it provides reasoning in a domain with a small body of knowledge [34]. Previous research has reported that ABE is able to produce more successful results in comparison to traditional regression based methods [3][22]. ABE has been favored over other methods when the dataset contains discontinuities [11]. However, it was remarked that ABE method is subject to a variety of decisions that have a strong impact on its predictive performance. Such decisions include selection of features and/or instances, deciding on the number of analogies to be used and the adaptation strategy [15][19]. Kocaguneli et al. [15] stated that using different solutions for each parameter produce different ABE configuration, hence, different ABE models. Therefore, there is a reasonable belief that choosing the right ABE model is not an easy process. One option is to study the characteristics of a dataset and come up with the suitable choice for each decision. We can reach to more manageable set of ABE models if researchers critically review the space of options for their models.

Indeed, there is a direct evidence that the choice of right adaptation method has a big influence on the accuracy of ABE as confirmed in [3][13]. Basically, the adaptation method of ABE is composed of three interrelated decision variables: (1) number of nearest analogies, (2) nominated set of features and (3) adaptation strategy weights. The purpose of this process is to generate more accurate estimates and minimize the difference between the estimated effort and actual effort. The original ABE method [22], that is denoted as ABE0, does not use any kind of adaptation strategy, but it uses the mean of $k$ nearest neighbors' efforts. The $k$ value here is determined manually by an expert for which the overall performance of the whole dataset is improved, but not necessarily the best performance for each individual project. The key challenge here is that the experts tend to find the optimum $k$ value based on minimizing one evaluation measure ignoring other evaluation measures, whereas the final model is evaluated using multiple evaluation measures. Previous studies showed that applying different evaluation measures tend to behave differently in identifying best model [32], therefore finding these decisions should be based on improving all evaluation measures simultaneously. Moreover, the improved ABE models that use adaptation strategy such as regression towards the mean [10], genetics algorithm [5] and neural networks [17] still fail in specifying the appropriate number of the nearest analogies and do not take other decisions in their adaption process.

Above all, we believe that finding the right decisions for the ABE adaptation method is a multi-objectives optimization problem. Therefore, in the present study we use Multi-Objective Particle Swarm Optimization (MOPSO) algorithm to tune and adapt ABE. The PSO is a relatively the most common used optimization method among researchers. It has been proposed by Kennedy and Eberhart [39] to perform combination of random and neighborhood search. It mimics the process of birds in searching for foods, further details about PSO can be found in Section 3. However, the conventional PSO can deal with problems that have only one objective function, but when the problem has many conflicting objectives as in our study we should use the extended version of PSO that can support multi-objective functions which is called multi-objective Particle Swarm Optimization (MOPSO) [37][40]. The goal of MOPSO is to find a set of solutions (called Pareto front) that improves the performance of two or more objective functions possibly subject to some constraints on the independent variable ranges. The objective functions used in this study are the evaluation measures such as Standardized Accuracy (*SA*), Mean Inverse Balanced Relative Error (*MIBRE*), and Mean Balanced Relative

Error (*MBRE*). Since these measures tend to behave differently [32], the final outcome of MOPSO is not a single solution but a set of solutions that make a good trade-off between these objective functions. In this study, each possible solution is composed of three variables: (1) number of nearest analogies (*k*), (2) set of nominated features and (3) adaptation strategy weights.

This paper is an extension to our previous works on using optimization for ABE adaptation [44]. In that work we only evaluated the local tuning on adjusting ABE without studying the impact of features and their optimized weight. Bearing that in mind, this paper aims at answering the following research questions:

-*RQ1*. Is there sufficient evidence that the MOPSO algorithm can find the best *k* value for each project individually?

-*RQ2*. Is there significant difference between Local adaptation and Global adaptation and which of them can improve overall performance for one project? Local adaptation means finding for each project the appropriate decisions that improve its individual accuracy, whereas Global adaptation means finding one common decision for all projects that improve accuracy for the whole dataset on average. Previous studies suggest that adapting each project individually with its own decisions tend to be more accurate than adapting whole dataset with the same decision vector [29].

-*RQ3*. Does the use of all features within the proposed adaptation strategy have rooms for further improvement rather than using optimized features set?

-*RQ4*. Does the use of equal weights produce more accurate results than the optimized weights in the adaption strategy?

The paper is structured as follows: Section 2 introduces an overview to ABE and adaptation methods. Section 3 presents the problem, where Section 4 introduces the objective functions and evaluation measures that will be used during optimization process. Section 5 introduces the MOPSO and its fundamental concepts. Section 6 presents the research methodology. Section 7 shows the descriptions of all datasets used in this study. Section 8 presents experimental setup. Sections 9, 10, 11 and 12 present answers to RQ1, RQ2, RQ3 and RQ4, respectively. Section 13 shows further analysis and comparisons between the proposed adaptation strategy and other adaptation methods. Finally, the paper ends with Section 14 which summarizes our work and outlines main conclusions.

## 2. *Background & Related Work*

ABE generates a new estimation based on assumption that similar projects with respect to features description have similar efforts [22]. The basic procedure of the ABE method is illustrated in Figure 1 where in the first step the training projects are preprocessed which includes data scaling, handling missing values and performing feature selection if necessary. The second step is to define a new project to be estimated. Then retrieve the similar projects that have been encountered and remembered as successful historical projects using a similarity measure such as Euclidean distance as shown in Eq. 1. Finally, the retrieved solutions are adapted and calibrated to minimize difference with the new project. We refer to the baseline ABE method that does not include adaptation as ABE0.

$$dis\tan ce = \sqrt{\sum_{c=1}^{m}(p_{ic} - p_{jc})^2} \qquad (1)$$

where *m* is the number of features, $p_i$ and $p_j$ are projects under investigation:

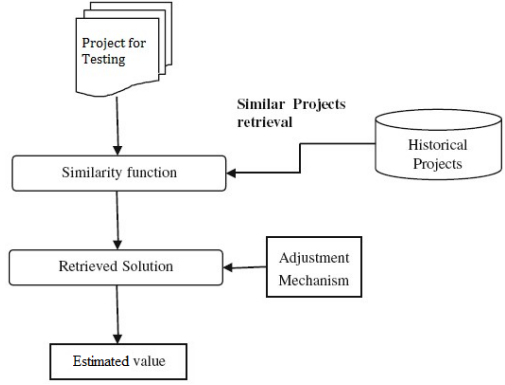

Fig. 1 Process of ABE method

This paper focuses mainly on the adaptation process which is known as local tuning of ABE method. To address the local tuning problems of ABE method, we have to better understand the relationship among three interrelated key factors of any successful adaptation method. These factors are: a) optimum features set, b) number of nearest analogies (*k*) and c) The weighting schema used in adaptation. By combining these factors together, we can easily find over thousands of different adaptation models. To select the right model, one has to try many options from the large space of configuration possibilities and then choose what performs better on the local data. This process is daunting and time consuming and hardly to be completed due to the diverse behavior of evaluation measures. Therefore, we suggest to treat this problem as multi-objective optimization problem in which the best solution can be found after searching on a large space of available solutions. We believe that the best solution should make trade-off between various evaluation measures since changing the value of any decision variable tends to behave diversely. Researchers when encounter the literature, they can find many studies that handle this problem but separately, in other words, some studies focus on the problem of predicting the best *k* number of nearest analogies, while others focus on the weighting methods of adaptations. So we cannot find any study that treat all three interrelated decision variables together in one model.

The use of adaptation strategy with ABE has seen significant improvements in terms of accuracy and reliability. Adaptation is a process that attempts to minimize the difference between test observation and each nearest observation and reflects that difference on the derived solution in order to obtain better accuracy. Then all adapted solutions are aggregated using either simple statistical approach such as mean, median or Inverse Ranked Weighted Mean (IRWM) as shown in Eq. (1) and (2), or by more sophisticated approaches such as machine learning algorithms.

$$\bar{e}_t = \frac{1}{k}\sum_{i=1}^{k}\hat{e}_i \qquad (1)$$

$$\bar{e}_t = \frac{\sum_{i=1}^{k}(k+1-r_i)\times \hat{e}_i}{\sum_{i=1}^{k}i} \qquad (2)$$

Where $\bar{e}_t$ is the new estimated effort and $\hat{e}_i$ is the adapted effort of the nearest $i^{th}$ analogy.

Before we start reviewing the existing adjustment method we should first mention the basic approach of null EBA adjustment which is based on finding the average of efforts for the nearest k projects. In late 1990s, Walkerden and Jeffery [24] introduced the first adjustment method called Linear Size Extrapolation (LSE). This method aims to calibrate the effort of a new project by making extrapolation between the size of the new project and the size of the nearest project. The principal reason of using only size feature was twofold: (1) it carries useful information about the project such as the amount of software functionality, (2) it has strong statistical significant correlation with project effort. Although the predictive performance of this method was notably superior to null adjustment over very limited number of datasets, there was criticism about the performance of this approach since not all datasets present strong correlation between effort and size feature. Thus, the prediction obtained will possibly be far from optimum. Furthermore, some datasets are described with number of different size features as in the case of web projects, therefore using only one size feature is not entirely reasonable.

Based on the above limitations, Kirsopp et al. [12] extended Walkerden and Jeffery method to include all size related features. This method is called thereafter Multiple Linear Feature Extrapolation (MLFE). The performance of MLFE has been reported in the study conducted by Mendes et al. [19] based on a dataset collected from web projects. Unfortunately, the replication study on adjustment methods [3] revealed that MLFE is still less useful than LSE under certain experimental conditions. The certain limitations of this approach are: it does not support categorical features and may not work well when any used historical feature contains zero as a denominator, which leads to infinity. The classical solution conducted is to ignore features with zero values.

On the other hand, an investigation analysis of many software projects reveals that expert judgment approach is considerably "productivity based adjustment". In this context, Jorgensen et al. [10] proposed a different method called Regression Towards the Mean (RTM) to adjust and calibrate nearest projects based on the notion of project productivity. This method assumes that if the nearest projects have extreme productivity values, then the productivity value of the project under estimation should be tuned to bring it closer to the average projects' productivity values in the dataset. Jorgensen et al. [10] remarked that the productivity distribution of estimated projects are narrower than that of actual projects which proves that the estimated efforts regress towards the mean effort in a particular dataset. Shepperd and Cartwright [23] replicated the work of Jorgensen et al. [10] and advised that the dataset should be partitioned into groups of homogeneous projects so that the adjustment moves to a local productivity mean. Through evaluation of some datasets, there was significant improvement on the accuracy when RTM method is applied.

Indeed, all previous adjustment methods rely primarily on the project size in order to adjust and calibrate new projects. However, this is not the only solution existing in literature. Li et al. [16] demonstrate that the similarity degree between projects can play important role in adjusting selected projects. Similarity based adjustment appears as a well suited method because it reflects the amount of differences between new project and its nearest analogies on the predicted effort. However, similarity measure can be divided into two levels: local measure and global measure. Local similarity measure is used to find out the similarity degree between two projects with respect to a particular feature where Global similarity measure is used to aggregate all local similarity values. In this approach, the local similarity degrees or global similarity degrees can be used to adjust the selected project by applying sum of product of the normalized similarity and effort. The main advantage of this approach is that it supports both categorical and continuous features.

Previous analysis studies report that software datasets are characteristically noisy with complex structure [29]. Therefore the statistical methods cannot learn the differences from the structure of selected projects. Genetic algorithm is used as potential solution to minimize differences between the target project and selected projects. Genetic Algorithm (GA) is a search heuristic method that simulates the natural evolution process. It is widely used in software engineering to solve complex problems such as those encountered in testing and management. A typical example on this approach is the work of Chiu and Heung [5] who used GA to calibrate selected projects based on learning distances between them and reflect that difference on the predicted effort. The adjustment function uses GA to find the optimized value for the adjustment coefficient $\alpha_j$ based on minimizing one of error measure. The main challenge when using GA is that it needs careful setup for its parameters such as specifying how to encode chromosome, how to perform mutation and crossover, and so forth.

Likewise, Azzeh [2] used Model Tree to adjust and tune selected projects. Model Tree is a type of decision tree model designed for non-linear regression, where the leave nodes have regression functions instead of numerical values. This adjustment method consists of two stages: Learning and prediction. During the learning phase, the differences between each historical project and its nearest project in the training dataset are computed across all features including efforts. These differences are then used to construct a Model Tree where differences in effort values are considered the output, and differences with respect to features are considered inputs. During the prediction phase, the nearest project to the new project is identified, then differences between them across all features are entered to the constructed Model Tree in order to compute the possible difference in the effort. This amount of difference is then added to selected effort to produce hopefully better estimate.

On the other hand, Li at al. [17] raised an important concern regarding structure of datasets as they claim that most software cost estimation datasets do not follow uniform distribution as presumed in linear methods. They used artificial neural networks, specifically Multilayer Perceptron, to learn difference in effort from differences in other features. The neurons are arranged in layers where the input data (i.e. feature distances between each source project and its analogy in the same historical dataset) are fed to the network at the input layer. The data then passes through hidden and output layers to produce the solution for a given problem. The solution here is the distance between projects effort. The learned differences are then added to the selected effort. The findings from the Li et al study are promising, but the replication study conducted by Azzeh [3] reported discouraging results where some linear adjustment methods produced more accurate results than neural networks. One possible explanation for these contradictory results is the fact that Multilayer Perceptron can be very sensitive to parameters choice in software effort estimation. So, they can perform very differently depending on the parameters choice.

From these adaptation methods, we can notice that there is no consensus regarding the use of features. Each method uses different set of features, for example: LSE used only the size feature, MLFE used a set of size features, RTM used size and effort features, GA and NN used all features. Likewise, there is no consensus regarding the best number of nearest analogies ($k$). In recent years, various approaches have been proposed to specify this number such as $k$ nearest neighbor algorithms and similarity cut off point [16][22]. Some studies favor using a pre-determined number of analogies in software effort estimation studies such as $k=1, 2… n-1$ [1][2][11][15][19][24]. In this approach, the practitioner starts with $k=1$ and increase this number until no further improvement on the accuracy can be achieved. Examples on this approach are the studies made by Lipowezky et al. [18] and Walkerden and Jeffery [24] who found $k=1$ was the most optimum number. Mendes el al. [19] used $k = 1, 2, 3$ as optimum numbers. Lipowezky et al. [18] proposes a policy that looks for only one prototype, which can be regarded as extreme when dealing with datasets as small as those in software effort estimation. Furthermore, we would like to base our estimations on some sample set of past data, not only on one record, since only one record may be misleading in small and heterogeneous datasets. On the other hand, Azzeh [3] conducted an extensive replication study on various linear and non-linear adaptation strategies using many public datasets, and found that that $k=1$ was the most prominent number across all experimentations. Kirsopp et al. [12] on the other hand proposes making predictions from the 2 nearest cases as it was found as the optimum value for the datasets of their study. In a further study Kirsopp et al. [19] have increased their accuracy values with case and feature subset selection strategies.

Besides this approach, other researchers attempted to dynamically find the optimum number of nearest analogies such as [16] and [29]. Li et al. [16] proposed a method to learn the $k$ number based by optimizing similarity threshold. They conducted extensive experimentation on actual and artificial datasets and observed various effects of $k$ values. Azzeh and Elsheikh [29] utilized bisecting k-medoid clustering to understand the structure of certain dataset and come up with best number of analogies for each test project individually. This method is somehow different than other methods because the authors proposed to make dynamic selection for each test instance in the dataset rather than using fixed $k$ values for the whole dataset. The results obtained from this study were promising and still needs further development. Above all, we still believe that the best $k$ number can be found through optimization algorithm which forms motivation for this research.

Above all, the results from adaptation studies are usually controversial and cannot be generalized since there are many uncontrollable sources of variations between these studies. In spite of their good performance for ABE method, most adaptation methods are still long way from reaching the real optimal solutions. This can be observed from the inconsistent behavior of such methods. In other words, they perform well in most cases over specific datasets and worse for the remaining. Some of them focus on one side of the problem ignoring the other sides which make producing high accuracy is relatively impossible. A challenging approach to address this issue is to study the impact of dataset characteristics in further details and the relevant of adaptation for each dataset. Therefore, to better understand the problem of adaptation, taking into consideration all sides of the problem, we propose to use multi-objective optimization to make better adaptation for the retrieved analogies. The main theme of this research is not only running individual studies but developing a better understanding of how beneficial is the proposed approach for software industry and at the very end consolidating a body of knowledge. It also allows us to derive lessons from using these techniques for better model performance. If the results are compatible, they can be considered additive, increasing confidence in the original hypothesis.

## 3. Problem Representation

This section describes the proposed adaptation strategy that is used to tune and adapt nearest analogies. The proposed function of ABE adaptation is illustrated in Eq. 3 and 4. The functions are composed of three decision variables: 1) number of nearest analogies ($k$), 2) feature distance weights ($w$) and 3) feature set ($v$). Eq. 3 shows how each project is tuned whereas Eq. 4 is used to aggregate the adapted projects' efforts using Ordered Weighted Mean (OWM). In OWM, a method with rank ($k$-$i$+1) gets a weight $\left(\frac{2^{i-1}}{2^k - 1}\right)$, as shown in Eq. 4 where $k$ is the number of nearest analogies and $i$ is the rank of a nearest project. For example, using 3 projects, one might give weight (4/7) for the top ranked project ($\hat{e}_1$), (2/7) for the next one ($\hat{e}_2$) and (1/7) for the 3rd project ($\hat{e}_3$) so Eq. 4 would appear as follows: $\bar{e}_t = \left(\frac{4}{7} \times \hat{e}_1 + \frac{2}{7} \times \hat{e}_2 + \frac{1}{7} \times \hat{e}_3\right)$.

$$\hat{e}_i = e_i + \frac{1}{m} \sum_{j=1}^{m} w_j \times v_j \times (f_{tj} - f_{ij}) \qquad (3)$$

$$\bar{e}_t = \sum_{i=k}^{1} \left(\frac{2^{i-1}}{2^k - 1}\right) \times \hat{e}_{k-i+1} \qquad (4)$$

Where $e_i$ is the effort of the nearest $i^{th}$ analogy. Based on the above assumption, each possible solution ($\vec{x}$) in the search space is represented as a vector of three decision variables as shown in Eq. 5.

$$\vec{x} = \langle k, v, w \rangle \qquad (5)$$

Where $k$ is the value that represents the number of nearest analogies which should be bounded by minimum of 1 and maximum of the size of training projects (i.e. $k \in [1, n]$ where $n$ is the number of training projects). $v$ is a binary vector whose coordinate represent the presence or absence of feature in the adaptation function. For simplicity use with MOPSO, we use $v$ as integer number instead of set of binary values and then we convert that value into its corresponding binary numbers when it is applied to Eq. 3. So the possible range for $v$ as integer would lie between *1* and *$2^m$-1* where $m$ is the maximum number of features in the training dataset. $w$ is a dynamic matrix of size ($n \times m$) as shown in Figure 2, which contains the possible weights to tune feature distance as depicted in Eq. 3. Each possible weight would take value between zero and one (i.e. $w_{ij} \in [0, 1]$) and the summation of weight values along a particular row should equal 1 as shown in Eq. 6.

$$\sum_{j=1}^{m} w_{ij} = 1, \forall i = 1,2,\ldots,n \qquad (6)$$

$$w = \begin{bmatrix} w_{11} & w_{12} & \cdots & w_{1m} \\ w_{21} & w_{22} & \cdots & w_{2m} \\ \cdots & \cdots & \cdots & \cdots \\ w_{n1} & w_{n2} & \cdots & w_{nm} \end{bmatrix}$$

Fig. 2 Weight Matrix for one solution in the search space

To illustrate that, consider the following hypothetical solution vector $\vec{x} = \langle 4, 15, w \rangle$ and assume $n$=5, $m$=6. The $w$ matrix is given in Figure 3. This solution vector shows that only 4 nearest analogies are considered in Eq. 4, the possible feature set after converting $v$ to binary number is {0, 0, 1, 1, 1, 1} which means that all features are included except the first two features in the training dataset. It is important to note that the number of bits equals to $m$ and the first position from the left is the first feature in training dataset.

$$w = \begin{bmatrix} 0.2 & 0.15 & 0.05 & 0.1 & 0.4 & 0.1 \\ 0.4 & 0.3 & 0.1 & 0 & 0.15 & 0.05 \\ 0.2 & 0.5 & 0.05 & 0.05 & 0.05 & 0.15 \\ 0 & 0.1 & 0.25 & 0.15 & 0.3 & 0.2 \\ 0.1 & 0.15 & 0.05 & 0.2 & 0 & 0.5 \end{bmatrix}$$

Fig. 3 Weight Matrix with values for one solution in the search space

## 4. Objective functions

This section describes the evaluation measures that will act as objective functions during optimization process and be used later to evaluate the software effort estimation models. Evaluation measures typically comment on the success of a prediction model. The cornerstone of the evaluation measures is the absolute error (*AE*) between the actual effort of a particular project ($e_i$) and the predicted effort of that project ($\bar{e}_i$) as shown in Eq. 7. This measure should be as small as possible because large deviation will have opposite effect on the development progress of the new software project.

$$AE_i = |e_i - \bar{e}_i| \qquad (7)$$

Based on this key measure the researchers found a lot of evaluation measures that can work well in evaluating effort prediction models such as Magnitude Relative Error (*MRE*) and their aggregated forms such as Mean Magnitude Relative Error (*MMRE*) and Median Magnitude Relative Error (*MdMRE*). Recent studies arise important concerns about using *MRE* because it is unbalanced and yields asymmetry distribution [6][21][30]. Therefore it is not always reliable to use *MRE* and their derived measures to compare between prediction models or to evaluate a single model.

In this paper we used the Standardized Accuracy (*SA*) measure that has been proposed by Shepperd and MacDonell [30] as shown in Eq. 8 which is based on Mean Absolute Error (*MAE*) because it is not a ratio and it does not present asymmetric distribution as in *MMRE*. The *SA* is mainly used to test whether the prediction model in hand really outperforms a baseline of random guessing and generates meaningful predictions. So, the *SA* is important to test the reliability of the prediction model because it can be interpreted as the ratio of how much better a given model is than random guessing, giving a very good idea of how well the approach does. The Mean Absolute Error of random guessing ($\overline{MAE_{p_o}}$) is obtained as the mean value of a large number runs of random guessing. This is defined as, predict a $\bar{e}_i$ for the target case *i* by randomly sampling (with equal probability) over all the remaining *n - 1* cases and take $\bar{e}_i = e_r$ where *r* is drawn randomly from $1…n \wedge r \neq i$. This randomization procedure is robust since it makes no assumptions and requires no knowledge concerning population.

$$SA = 1 - \frac{MAE}{\overline{MAE_{p_o}}} \qquad (8)$$

In addition to the above mentioned evaluation measures, we used other three reliable evaluation measures mentioned in literature that are considerably less vulnerable to bias or asymmetry distribution as in case of *MMRE* [31][32]. These measures are Balanced Relative error (*BRE*) and the Inverted Balanced Relative Error (*IBRE*) and their average measures mean of *BRE* (*MBRE*) and the mean of *IBRE* (*MIBRE*) as shown in Eqs. 9, 10, 11, and 12 respectively.

$$BRE_i = \frac{AE_i}{min(e_i, \bar{e}_i)} \qquad (9)$$

$$IBRE_i = \frac{AE_i}{max(e_i, \bar{e}_i)} \qquad (10)$$

$$MBRE = \frac{1}{n}\sum_{i=1}^{n} BRE_i \qquad (11)$$

$$MIBRE = \frac{1}{n}\sum_{i=1}^{n} IBRE_i \qquad (12)$$

Where *n* is the number of projects in the dataset.

All evaluation measures are objectives to be minimized except *SA* which is to be maximized. These measures were chosen because, even though all of them were initially designed to represent how well a model performs, they can behave very differently from each other as reported in [32]. This allows us to select as many as possible good solutions that can make trade-off between these measures.

## 5. Multi-Objective Particle-Swarm Optimization algorithm
### 5.1 Basic Concepts

Optimization algorithm is a typical solution for the sophisticated problems that have many interrelated design options as encountered in software engineering tasks [5][32]. The problem of optimization can be defined as follows: Given a function $f: X \rightarrow \Re$ from some set of decision vectors (*X*) to the set of real numbers ($\Re$), the aim is to find a solution $\vec{x}_o$ in *X* such that the objective function is either minimized ($f(\vec{x}_o) \leq f(\vec{x}), \forall \vec{x} \text{ in } X$) or maximized ($f(\vec{x}_o) \geq f(\vec{x}), \forall \vec{x} \text{ in } X$), where each solution $\vec{x}$ can be defined as vector of decision variables in the *m*-dimensional space as shown in Eq. 13.

$$\vec{x} = [x_1, x_2, \ldots, x_s]^T \tag{13}$$

Many problems can be solved based on optimizing a single objective function, but when the problem has many objectives and these objectives are in conflict with each other, we come to situation that there is no single solution but instead there is a good trade-off solutions that represent best compromises among the objectives. The problem of ABE adaptation procedure can be viewed as multi-objective optimization algorithm which contains many interrelated decisions that need to be optimized based on finding trade-off between evaluation measures. As there are many optimization algorithms in literature, we chose Particle Swarm Optimization (PSO) for two reasons: (1) The algorithm is simple and its implementation is straightforward, (2) it showed good performance against some well-known evolutionary algorithms such as Genetics algorithm and Simulated Annealing [36][37]. In the next sections we provide an introduction to PSO algorithm and its application to multi-objectives problems.

### 5.2 Particle Swarm Optimization

The PSO is a population based heuristic search algorithm which simulates the movements of a flock of birds to find food. It was first developed in 1995 by Kennedy and Eberhart [39]. Basically, the algorithm performs a kind of local and global search combined with random search. This algorithm was originally proposed for balancing weights in neural networks, then soon later became one of the best optimization algorithms. The popularity of PSO stems from its simplicity in performing search and especially global search since it does not need many operators for creating new solution as in evolutionary algorithms, so its implementation is straightforward [37]. But on the other hand, this algorithm suffers from two main problems: 1) slow convergence in refined search stage, and 2) Weak local search ability [40]. These two problems are handled when using different search topology as mentioned later in this section. Before we demonstrate how PSO works we list the key concepts of PSO in Table 1.

TABLE 1. Key concepts of PSO algorithm

| Concept | Meaning |
| --- | --- |
| *Swarm (X)* | It is the population of the algorithm which contains number of Particles. |
| *Particle ($\vec{x}$)* | Represents the potential solution as vector of *M* decision variables in the swarm. |
| *pbest* | It is the best position of a Particle that has been achieved so far. |
| *gbest* | It is the global position of the best particle in the swarm. |
| *leader* | Represent the Particle that is used to guide other particles. |
| *Velocity vector ($\vec{V}$)* | It derives the optimization and determines the direction to the next move. |
| *Inertia weight (W)* | It is used to control the impact of previous velocities on the current Particle's velocity. |
| *Learning factor ($C_1$ and $C_2$)* | Represent the attraction of a particle to its own success or that of its neighbors. |
| *Neighborhood topology* | It specifies the structure of the swarm and how the Particles are connected. |

The algorithm starts with population initialization of random solutions and velocities, and then searches for optima by updating the generations. Particles then fly through the problem space by following the current optimum Particles [39]. The position of a Particle is changed according to its own flying experience as well as the flying experience of neighbours. The *pbest* and *gbest* are updated accordingly. To illustrate that, consider that $\vec{x}_i(t)$ is the position of the $i^{th}$ Particle at time *t*. The new position of that Particle is updated by adding the amount of velocity ($\vec{V}_i(t)$) on its previous position as shown in Eq. 14 and 15. Moreover, all particles in PSO are kept as members of the population through the course of the run [38].

$$\vec{V}_{ij}(t) = W \times \vec{V}_{ij}(t-1) + C_1 \times r_1 \times \left(\vec{x}_{pbest_i} - \vec{x}_{ij}(t)\right) + C_2 \times r_2 \times \left(\vec{x}_{leader} - \vec{x}_{ij}(t)\right) \tag{14}$$

$$\vec{x}_{ij}(t) = \vec{x}_{ij}(t-1) + \vec{V}_{ij}(t) \tag{15}$$

Where $r_1$ and $r_2$ are random values [0, 1], *j* represents the index of the decision variable in $\vec{x}_i(t)$.

It is interesting to note that a large inertia weight (*W*) facilitates a global search while a small inertia weight facilitates a local search [39]. By linearly decreasing the inertia weight from a relatively large value to a small value through the course of the PSO run gives the best PSO performance compared with fixed inertia weight settings. The success of Particle move depends on the success of the other connected particles which are not necessarily be the Particles that are close to each other, but instead the Particles that are close to each other based on the neighborhood topology that defines the swarm structure. Particles can be connected to each other by different topologies represented as a graph, typical examples on PSO topologies include: 1) Empty graph, 2) Local Best, 3) Fully connected Graph, 4) Star network, and 5) Tree network. Each one of these topologies has different implementation and considerations and thus leads to different performance. Fully Connected graph is the widely used topology because it allows fast convergence than others.

However, the classical PSO can deal efficiently when the problem has only one objective function, but when the problem has many conflicting objectives, we should use the extended version of PSO that can support multi-objective functions which is called multi-objective Particle Swarm Optimization (MOPSO) [40] as explained in the next section.

### 5.3 Multi-Objective Particle Swarm Optimization (MOPSO)

The MOSPO [40] is concerned with the problems that consists of one or more decisions and have many objectives to be optimized simultaneously. The problem of multi-objective optimization can be defined as finding a vector of design options or decisions ($\vec{x}$) which will satisfy *d* inequality constraints ($g_i(x) \geq 0, \quad i = 1, 2, \ldots, d.$) and *p* equality constraints ($h_i(x) = 0, \quad i = 1, 2, \ldots, p.$) and simultaneously optimizes a vector of *M* conflict objective functions as shown in Eq. 16. The constraints mentioned above define the feasible region which includes all the admissible solutions. The optimal solution is obtained as trade-off between two or more conflicting objectives which is called Pareto optimal solution. Below are some important definitions pertaining to the multi-objective optimization:

$$\vec{f}(\vec{x}) = [f_1(\vec{x}), f_2(\vec{x}), \ldots, f_M(\vec{x})] \tag{16}$$

**Definition 1** (Dominance): A solution $\vec{x}_i \in \Re^m$ is strictly dominated by a solution $\vec{x}_j \in \Re^m$ ($\vec{x}_i \prec \vec{x}_j$ and $\vec{x}_i \neq \vec{x}_j$) if and only if $f_l(\vec{x}_i) \leq f_l(\vec{x}_j), \quad \forall l \in 1, 2, \ldots, M$ and $f_l(\vec{x}_i) < f_l(\vec{x}_j), \quad \exists l \in 1, 2, \ldots, M$. This concept can be easily extended to maximization problem.

**Definition 2** (Non-Dominance): The solution $\vec{x}_i \in \Re^m$ is called non-dominated solution, if none of the objective functions can be strictly improved in value without degrading some of the other objective values. In other words, it is non-dominated if there does not exist another solution $\vec{x}_j \in \Re^m$ such that $f_l(\vec{x}_i) \prec f_l(\vec{x}_j), \quad \forall l \in 1, 2, \ldots, M.$ and $i \neq j$

**Definition 3** (Pareto Optimal): We say that a solution $\vec{x}^* \in \eta \subset \Re^m$ is a Pareto optimal if it is non-dominated with respect to the feasible region ($\eta$)

**Definition 4** (Pareto Optimal set): a set $\rho' \subset X$ of non-dominated solutions is called Pareto Optimal set which is formally defined as: $\rho = \{\vec{x} \in \eta \mid \vec{x} \text{ is a Pareto optimal}\}$

**Definition 5** (Pareto Front): is defined as $\rho f = \{f(\vec{x}) \in \Re^m \mid \vec{x} \in \eta\}$

The key question when extending the PSO to support Multi-Objective problems is how to find global best solution *gbest* that acts as a leader and guide for other Particles. The answer is very important because it affects both the convergence to the true Pareto front and a well-distributed front. Since there is no single best solution in Multi-Objective problems, the non-dominated solutions found by MOPSO are often stored in a special repository where each Particle can select randomly a non-dominated solution from that repository to act as global guide for its new Position. In spite of its simplicity, it cannot promote convergence. In this research, we used the more efficient MOPSO algorithm based on Crowding Distance (MOPSO-CD) [40]. The crowding distance factor (*CD*) gives an estimate of the density of solutions surrounding a non-dominated solution and show how much a non-dominated solution is crowded with other solutions. So instead of randomly selecting *gbest* from the whole solutions in the repository, it is randomly selected from the top 10% less crowded area of the repository for each Particle that is dominated by any solution located in this area [40] [41].

Figure 4 describes the algorithm of complete MOPSO-CD. At the first step, the swarm is randomly initialized with a predefined number of Particles and its initial velocities. In this step, the content of each Particle is generated randomly and the initial velocities are preferably set to zero. Also, the *pbest* of each particle is set to its initial position. This step is fully described in Figure 6. In the second step, The fitness of all Particles in the swarm are evaluated based on their current position. The non-dominated Particles are then selected and stored in a special repository (*A*) [41]. In each iteration the repository is updated and new non-dominated Particles are added. The Particles in the repository are then evaluted against the new added Particles and the dominated ones are deleted. It is important to note that the capacity of the repository is limited so when the number of non-dominated Particles exceeds maximum capacity then it is reduced based on applying *CD* factor. The *CD* is calculated for non-dominated solution by first sorting the solutions in ascending order according to each objective function [41]. For each objective function, the crowding distance for each Particle is calculated by finding the distance of its neighbors as shown in Eq 17. The *CD* factor of first and last solution is usually equal to the maximum distance [41]. The final *CD* for each Particle is the sum of *CDs* along all objective functions. Figure 5 describes the pesudo code of obtaining *CD* factors.

$$CD_i = f_{i+1} - f_{i-1} \tag{17}$$

The Particles with less crowded spaces are kept whereas the Particles with smallest *CD* factors are strictly deleted. Likewise, the value of *pbest* solution for each Particle is examined against current solution and the new *pbest* is determined based on one of three ways [40]: 1) if the *pbest* is dominated by current solution then the current solution is the new *pbest* for that Particle. 2) if the current solution is dominated by *pbest* then nothing changed, 3) Otherwise, one of them is selected by random as *pbest*. Meanwhile, the velocity of each particle is updated by using Eq. 14. It is interesting to note that the *gbest* in MOSPOS-CD represents a solution that is being randomly selected form the repository with less 10% crowded solutions. The new position of each Particle is updated using Eq. 15, but every time the solution is being updated the boundary values of each dimension variable in the solution is checked and adapted as shown in Algorithm 4. The above procedure is repeated until the maximum number of iterations (*T*) is complete.

During the update of Particles, it is important to mutate the current solution [41]. The mutation procedure is a crucial task in MOPSO to prevent premature convergence due to existing local Pareto fronts in some optimization problems. The mutation procedure used here is straightforward which adjust the position of a Particle by either adding or subtracting a specific value (*y*) depending on both the current iteration and either

Upper Bound value (UB) and lower bound value (LB) of each dimension in the solution space as shown in Eqs. 18 and 19.

```
1:   input: pop_size, C_1, C_2, W
2:   for i=1 to pop_size
3:       x_i ← initialize()
4:       v_i ← initialize()
5:       f_i ← evaluateFitness(x_i)
6:       Pbest_i=x_i
7:       Gbest_i=BestParticle(X)
8:   end for
9:   t=0
10:  A ← Non-Dominated(X)
11:  while (t<T)
12:      foreach x in A
13:          CD_i ← CrowdingDistance(x)
14:      end foreach
15:      A ← sortByCD(A)
16:      for j=1 to pop_size
17:          gbest_i ← Select-Global-Guide-Randomly-From-Top10%(A)
18:          v_j[t+1] ← updateVelcoity(v_j[t], pbest_j, gbest_j)
19:          x_j[t+1] = x_j[t] + v_j[t+1]
20:          if (x_j[t+1]> MaxBoundery or x_j[t+1]< MinBoundery)
21:              v_j[t+1]=-1× v_j[t+1];
22:              x_j[t+1]= x_j[t+1]+ v_j[t+1]
23:          end if
24:          if(t < T × Pmut)
25:              x_j[t+1] ← mutation(x_j[t+1]
26:          end if
27:          x_j[t+1] ← evaluateFitness(x_j[t+1])
28:      end for
29:      A ← A ∪ Non-Dominated(X)
30:      if(A is full)
31:          for each x in A
32:              CD(x) ← CrowdingDistance(x)
33:          end foreach
34:          A ← sortByCD(A)
35:          A ← A ∩ LessCrowdedSolution(A)
36:      end if
37:      x_j[t+1] ← Update(x_j[t+1])
38:      pbest_j ← Check(pbest_j, CurrentPosition)
39:      t++
40:  end while
```

Fig. 4 Pseudo Code of the complete MOSPO-CD

```
1:   S ← GetNumberOfNon-dominatedSol(A)
2:   for i=0 to S
3:       x_i.distance = 0
4:   end for
5:   for each Objective function M
6:       A ← sort(A, M)
7:       for i=1 to n-1
8:           x_i.distance = x_i.distance + (x_{i+1}.f_M − x_{i−1}.f_M)
```

```
9:      end for
10:     x_0.distance = x_S.distance = f_M^{max}
11: end for each
```
Fig. 5 Pseudo Code of CrowdingDistance()

$$x_{ij} = \begin{cases} x_{ij} + \Delta(t, UB_{ij} - x_{ij}) & if\ R = 0 \\ x_{ij} + \Delta(t, x_{ij} - LB_{ij}) & if\ R = 1 \end{cases} \quad (18)$$

$$\Delta(t, y) = y \times \left(1 - r^{1-\left(\frac{t}{MaxItr}\right)^b}\right) \quad (19)$$

Where:
- *R* is a randomly generated bit (zero and one both have a 50% probability of being generated)
- *t* is the current iteration number
- *r* is a random number generated from a uniform distribution in the range [0,1]
- *b* is a tunable parameter that defines the non-uniformity level of the operator. In this approach, the *b* parameter is set to 5 as suggested in [40].

## 6. Methodology
### 6.1 Using The Solutions Produced By MOPSO-CD in

The solutions generated by MOPSO-CD are used in the adaptation strategy of ABE in two ways: Local tuning and global tuning as will be discussed in section 6.2. These solutions are considered the best-fit Pareto solutions with the best train *MIBRE*, best train *MBRE* and best train *SA*. In order to show how the MOPSO-CD algorithm works with ABE we first start with describing the process of initialization as shown in Figure 6. The pseudo code shown how the Particles are initialized with their initial velocity. As we have seen earlier that each Particle represent potential solution which is composed of three variables: *k*, *v* and *w*. Each Particles is initialized with random values for each variable. For example the value of *k* can take integer number between *1* and *n-1* where *n* is the number of projects in the dataset. The value of *v* is also initialized with random integer number between 1 and $2^m-1$. The value of *v* is converted into binary number during the main run to represent the presence or absence of features in the adaptation process. Finally, the matrix *w* is initialized with random numbers between *0* and *1* bearing in mind that the summation of each row vector must equal one. The initial velocity for each Particle is also initialized with value 0.

```
1:  \vec{x}_i = []
2:  \vec{x}_i.k = Rand(1, n-1)
3:  \vec{x}_i.v = Rand(1, 2^M - 1)
4:  for j=0 to n-1
5:    for d=0 to m-1
6:      \vec{x}_i.w_{jd} = Rand(0, 1)
7:    end for
8:  end for
9:  return \vec{x}_i
```
Fig. 6 Pseudo Code of Initialize (N, M) Algorithm

When Particles fly to find better solutions the velocity and position of the Particle is updated based on its experience and that of neighborhoods [37]. The Pseudo code in Figure 7 shows how each particle is updated in this work according to Eq. 14. Recent research papers demonstrates that the velocity usually tends to exceed to a large value, which results in solutions go beyond the boundaries of the search space. This is more likely to happen when a solution is far from *gbest* and *lbest*. The typical solution is to truncate the location at the exceeded boundary at this iteration and reflect the velocity in the boundary so that the particle moves away at the next generation [38]. This technique does not necessarily alter the direction of Particle, but permitting the particle to stay in the vicinity of the boundary [38]. However, it does limit the solution step size, thereby preventing further divergence of solutions and permits the Particle to remain close to the boundaries during the search process. To make *x* feasible, the solution is dragged back along its line of movement until it reaches the nearest boundary.

```
1:  input: x, V, pbest, lbest
2:  V_i(t).k = W × V_i(t-1).k + C_1 × r_1 × (x_pbest_i.k - x_i(t).k) + C_2 × r_2 × (x_gbest.k - x_i(t).k)
3:  if (V_i(t).k > V_max(k) or V_i(t).k < V_min(k))
4:     V_i(t).k = -V_i(t).k
5:  end if
6:  V_i(t).k = min(max(V_i(t).k, V_min(k)), V_max(k))
7:  if (V_i(t).v > V_max(v) or V_i(t).v < V_min(v))
8:     V_i(t).v = -V_i(t).v
9:  end if
10: V_i(t).v = min(max(V_i(t).v, V_min(v)), V_max(v))
11:
12: for j=0 to N-1
13:   for d=0 to M-1
14:     V_i(t).w_jd = W × V_i(t-1).w_jd + C_1 × r_1 × (x_pbest_i.w_jd - x_i(t).w_jd) + C_2 × r_2 × (x_gbest.w_jd - x_i(t).w_jd)
15:     if (V_i(t).w_jd < 0 or V_i(t).w_jd > 1)
16:        V_i(t).w_jd = -V_i(t).w_jd
17:     end if
18:     V_i(t).w_jd = min(max(V_i(t).w_jd, 0), 1)
19:   end for
20: end for
21: return V
```

Fig. 7 Pseudo Code of the velocity update algorithm

## 6.2 Local Tuning Vs. Global Tuning

The solutions produced by MOPSO-CD are used for ABE in two ways: Local tuning (LT) and Global tuning (GT). The main difference between them is that in LT each project is tuned solely with its own solution vector, i.e. with its own *k* value, features set and weight values. Whereas in GT all projects in the dataset share the same optimum solution. During both procedures, the goal is to increase accuracy and decrease error rate. One point that needs clarification at this stage is how to use the objective functions in both types of tuning. Since applying evaluation measures for one project is totally different than evaluating the whole dataset, for example, we cannot apply *SA* evaluation measure on single project because it needs the average of absolute errors for all projects. Therefore, we made a little modification on the type of accuracy used in both tuning ways. During LT the optimum solution is optimized based on minimizing *AE*, *BRE*, *IBRE* after running MOPSO-CD for each project individually. In this case, when a particular project comes to be predicted, the MOPSO-CD is invoked to come up with optimum solution for that project bearing in mind to minimize *AE*, *BRE*, *IBRE* evaluation measures. So for *n* projects the MOPSO-CD is invoked *n* times. The final outcome of this process is *Pareto front* solutions for each project in the dataset. This tends to improve mainly the performance of each individual project and then overall performance of the dataset.

On the other hand, during GT all projects have the same solution vectors, i.e. all instances share the same decision values. Unlike LT, every possible solution here is applied to all instances in the training datasets and the objective function is calculated using the aggregated evaluation measures *SA*, *MIBRE* and *MBRE*. In this case, the MOPSO-CD is applied once since the adapted ABE model is run inside MOPSO-CD. In each run one generated solution is evaluated over all projects in the dataset in attempt to increase *SA* and minimize *MIBRE* and *MBRE*. Finally, we end up with one optimum solution that fits the whole dataset and improve overall performance, not the individual performance. Since the Pareto front may contain many solutions so to select the best solution among them is done as follows: The solutions in Pareto front are ranked based on the three employed evaluation measures, in other words the solutions are ranked in terms of each evaluation measure then the accumulative ranking is obtained by measuring the average ranking. The solution with minimum average ranking is selected.

## 7. Datasets

In order to assess the performance of any model, it is necessary to validate such model over some historical datasets that exhibit different characteristics as shown in Table 2. Most of the methods in literature were tested on a single or a very limited number of datasets, thereby reducing the credibility of the proposed model [26]. To avoid this pitfall, we included 15 software effort datasets that come from different industrial sectors. Specifically, these datasets come from two different sources namely, PROMISE [4] and ISBSG [7]. PROMISE is a publically available data repository and it consists of datasets donated by various researchers around the world. The datasets come from this source are: desharnais, kemerer, albrecht, cocomo, maxwell, china, telecom and nasa datasets. In our study, we also wanted to see the performance of ensembles on homogeneous datasets as well. Therefore, we only selected homogeneous datasets that are as big as the smallest heterogeneous dataset in terms of instance number. Cocomo dataset enables the researchers to classify projects in terms of three different development modes: Organic, semi-detached and embedded [25]. Therefore we used development mode as our breakdown criteria in cocomo and took three homogeneous subsets of cocomo: cocomo_O, cocomo_E and cocomo_S. cocomo_O includes organic projects whereas cocomo_E includes embedded projects and finally cocomo_S includes semi-detached projects. For the desharnais dataset, the development center of projects was the breakdown criteria. Projects in desharnais are developed in one of the three different development type. Like cocomo dataset, we took three subsets of desharnais based on their development type: desharnais_L1, desharnais_L2 and desharnais_L3. This process has results in 15 datasets.

The other dataset comes from ISBSG data repository (release 10) [7] which is a large data repository consists of more than 4000 projects collected from different types of projects around the world. Since many projects have missing values only 505 projects with quality rating "A" are considered. 10 useful features were selected, nine of which are numerical features and one of which is categorical feature. Since this dataset is not publically available and in order to allow replication for our study, the used features from ISBSG are represented in Table 3. One caution should be beard in mind here that although the ISBSG guideline suggests to use their criteria to select projects and features, there is no agreement among researchers about the features they have to choose. So we used the criteria that already utilized in our previous research which can be found in [2][29][46].

The employed datasets typically contain a unique set of features that can be categorized according to four classes [26]: size features, development features, environment features and project data. Table 2 shows the descriptive statistics of such datasets. From these statistics we can conclude that datasets in the area of software effort estimation share relatively common characteristics. They often have a limited number of observations that are affected by multicollinearity and outliers. Notably, all datasets have positive skewness efforts that range from 1.78 to 4.36 which indicate that the effort of each dataset is not normally distributed and presents a challenge for developing accurate estimation model.

TABLE 2 Statistical properties of the employed dataset

| Dataset | Feature | Size | Effort Data | | | | | |
|---|---|---|---|---|---|---|---|---|
| | | | unit | min | max | mean | median | skew |
| albrecht | 7 | 24 | months | 1 | 105 | 22 | 12 | 2.2 |
| kemerer | 7 | 15 | months | 23.2 | 1107.3 | 219.2 | 130.3 | 2.76 |
| nasa | 3 | 18 | months | 5 | 138.3 | 49.47 | 26.5 | 0.57 |
| ISBSG | 10 | 505 | hours | 668 | 14938 | 2828.45 | 1634 | 2.1 |
| desharnais | 12 | 77 | hours | 546 | 23940 | 5046 | 3647 | 2.0 |
| desharnais_L1 | 11 | 44 | hours | 805 | 23940 | 5413 | 3993.5 | 2.3 |
| desharnais_L2 | 11 | 23 | hours | 1155 | 14973 | 5095.4 | 3437 | 1.15 |
| desharnais_L3 | 11 | 10 | hours | 546 | 5880 | 1684.5 | 1123.5 | 1.86 |
| cocomo | 17 | 63 | months | 6 | 11400 | 683 | 98 | 4.4 |
| cocomo_E | 17 | 28 | months | 9 | 11400 | 1153 | 354 | 3.4 |
| cocomo_O | 17 | 24 | months | 6 | 240 | 60 | 46 | 1.7 |
| cocomo_S | 17 | 11 | months | 5.9 | 6400 | 849.65 | 156 | 2.64 |

| china   | 18 | 499 | hours  | 26    | 54620 | 3921   | 1829   | 3.92 |
| maxwell | 27 | 62  | hours  | 583   | 63694 | 8223.2 | 5189.5 | 3.26 |
| telecom | 3  | 18  | months | 23.54 | 1115.5| 284.33 | 222.53 | 1.78 |

TABLE 3 Feature description of ISBSG dataset

| Feature | Description | Type |
|---|---|---|
| AFP | Adjusted Function Points | Numerical |
| INC | Count of input functions | Numerical |
| OUC | Count of output functions | Numerical |
| EQC | Count of enquiry functions | Numerical |
| FILE | Count of files | Numerical |
| INF | Count of interface functions | Numerical |
| ADD | Count of added function | Numerical |
| CHC | Count of changed Functions | Numerical |
| RSL | Resource Level | Categorical |
| Effort | Effort needed to accomplish Software project in hours | Numerical |

## 8. Experiment Setup

The data preprocessing is important task in any prediction model so that we correctly assess the performance of new model with historical models.

- *Missing Values*: all projects with missing values in any feature were excluded from the dataset. For example the desharnais dataset consists of 4 projects have missing values therefore these projects have been ignored from the dataset, which resulted in 77 complete projects.
- *Standardization*: Since the features in the datasets present different scales, all continuous features were scaled using min-max transformation as shown in Eq. 20. This step is very important to eliminate the impact from different feature types, and to have same influence degree. Please note that some existing adaption methods such as LSE and RTM cannot work well with standardization since some features may contain zero after standardization and it is highly likely to divide on zero. To avoid this pitfall we followed the same solution conducted by Kirsopp et al. [12] in that all features that would introduce a zero into the denominator (for a particular case) are excluded from the calculation of the adaptation.

$$z_i = \frac{z_i - min(Z)}{max(Z) - min(Z)} \qquad (20)$$

- *Unnecessary Dependent features*: Since this study is designed for effort estimation, all unnecessary after-the-event features such as duration or time are removed from the data before starting any experiment.
- *Similarity function*: we used normalized un-weighted Euclidean distance measure to retrieve nearest analogies as shown in Eq. 1. In case of categorical features, we make binary comparison between two values, i.e. 0 if two projects have same feature values and 1 otherwise.
- *Validation strategy*: The leave-one-out cross validation has been used to validate and compare between different models. Although some authors favored n-Fold cross validation. The principal reason for this selection, the leave-one-out cross validation has been used in deterministic procedure that can be exactly repeated by any other research with access to a particular dataset [45]. According to previous studies, the leave-one-out cross validation generates lower base estimates than n-Fold cross validation since the methods need to learn from fewer examples. Also, it generates higher variance estimates than n-Fold cross validation since leave-one-out cross validation conducts more tests [45]. In each run, one project is held out as test instance and the remaining projects as training set. The prediction model is developed on the training set while the test set is used to evaluate the model. The error measures are calculated for each test instance. This procedure is continued until all projects within dataset act as test projects. Moreover, the proposed adaptation functions are compared to some well-known adaptation strategies existing in the literature such as: LSE [24], RTM [10], AQUA [16] and GA [5] in addition to the ABE0.

- Performance measures: In addition to the performance measures that are mentioned in section 4 as objective functions we used effect size ($\Delta$) as shown in Eq. 21 to check whether the predictions of the model in hand are generated by chance, and to justify if there is large effect improvement over guessing since the statistical significance test alone is not so informative if both predictions models are significantly different. The value of $\Delta$ can be interpreted in terms of the categories of small (0.2), medium (0.5) and large (0.8) where value larger than or equal 0.5 is considered better [30]. Besides that, we used Logarithmic Standard Deviation (*LSD*) as shown in Eq. 22.

$$\Delta = \frac{MAE - \overline{MAE}_{p_o}}{SP_o} \tag{21}$$

$$LSD = \sqrt{\frac{\sum_{i=1}^{n}\left(\lambda_i + \frac{s^2}{2}\right)^2}{n-1}} \tag{22}$$

Where:
- $MAE$ is the mean absolute error of the prediction model.
- $\overline{MAE}_{p_o}$ is the mean value of a large number runs of random guessing.
- $SP_o$ is the sample standard deviation of the random guessing strategy.
- $s^2$ is an estimator of the variance of the residual $\lambda_i$, and $\lambda_i = ln(e_i) - ln(\overline{e}_i)$

Presenting results without statistical significance is not convincing therefore we use *win, tie, loss* algorithm [25] to compare the predictive performance of the variants of adaptation methods, see Figure 8. To do so, we first check if two methods *Method$_i$*; *Method$_j$* are statistically different according to the Wilcoxon sum rank test; otherwise, we increase *tie$_i$* and *tie$_j$*. If the error distributions are statistically different, we update *win$_i$*; *win$_j$* and *loss$_i$*; *loss$_j$*, after checking which one is better according to the performance measure at hand *E*. The performance measures used here are *MAE, MBRE, MIBRE, LSD* and *SA*. We used Wilcoxon sum rank test because the models errors are not normally distributed as confirmed by Kolmogorov–Smirnov test. Also, the *win-tie-loss* algorithm used Wilcoxon test in its procedure [25].

```
1:   winᵢ=0,tieᵢ=0,lossᵢ=0
2:   winⱼ=0,tieⱼ=0;lossⱼ=0
3:   if WILCOXON(AE(Methodᵢ), AE(Methodⱼ), 95) says they are the same
4:   then
5:       tieᵢ = tieᵢ + 1;
6:       tieⱼ = tieⱼ + 1;
7:   else
8:     if better(E(Methodᵢ), E(Methdⱼ)) then
9:         winᵢ = winᵢ + 1
10:        lossⱼ = lossⱼ + 1
11:    else
12:        winⱼ = winⱼ + 1
13:        lossᵢ = lossᵢ + 1
14:    end if
15:  end if
```

Fig. 8 Pseudo code for *win, tie, loss* calculation between method *Method$_i$* and *Method$_j$* based on performance measure *E* [25].

## 9. The Performance of MOPSO in finding Best k values

This section presents an analysis of the Pareto solutions with concentration on finding the best *k* nearest analogies for each project. It also presents the answer to RQ1. Usually, researchers tend to start with one analogy and increase this number, then they use the *k* value that produces the best overall accuracy of the whole dataset. However, this choice may produce overall best accuracy but does not necessarily provide the best accuracy for each individual project, and could not be the perfect choice for other datasets. Furthermore, the selection of *k* value usually makes the prediction model behaves differently based on the employed evaluation measures. The obtained *k* value was frequently found based on using only one evaluation measure, usually *MMRE*, whereas the final model is evaluated using different evaluation measures. To better understand the behavior of *k* value selection we formulate an empirical analysis using Local tuning method (LT) with only one decision variable (i.e. *k*) is optimized in the solution vector, while other decision variables feature set (*v*) and weight matrix (*w*) remain unchanged during optimization process. Specifically, the *v* contains ones along all its dimension (i.e. using all features) and weight matrix contains equal weights (i.e. $w_{ij}$=1) for all possible values. In this analysis the outcome is a set of the best *k* values, one for each project, obtained as trade-off based on optimizing three evaluation measures (*AR, BRE, IBRE*).

Figure 9 shows the bar chart of the best selected *k* numbers for each examined dataset. Since few datasets are sufficient to demonstrate the behaviors of *k* selection we chose only four datasets telecom, albrecht, desharnais and maxwell. For a dataset of size *n* training observations, the best *k* value can range from 1 to *n*. The x-axis represents the *k* nearest analogies, and y-axis represents the number of projects selected that *k* value. The variability of *k* values shows that the projects in the dataset tend to use different *k* nearest projects in making final estimates. It is worth nothing that there is no definite pattern for the process of *k* selection, and there is no clear evidence that few or large number of analogies are sufficient to produce better estimates. This arise another important concern about validity of previous variants of ABE that relies heavily on expert intuition. Previous research studies [1][2][11][15][19][24] use limited number of analogies which is frequently less than or equal to 5 analogies. In spite of the limited efficiency of this mechanism it could ignore other useful analogies that can help to increase the accuracy. Therefore we believe that the expert intuition should be integrated with an automated procedure to predict the best *k* value.

Another important issue that we should focus in this analysis is the ability of MOPSO to generate not only one solution, but Pareto front solutions. The Pareto front contains various solutions that are not emphasizing a particular evaluation measure but on different evaluation measures. Minku et al. [32] showed that using different evaluation measures behave differently therefore they could be useful to produce prediction models that present trade-off between these evaluation measures. If the estimator has no reason for emphasizing a certain evaluation measure, s/he can analyze these solutions and use whichever solution s/he most prefers. If not, s/he may choose the solution more likely to perform best for that measure. At this case the solution that may appear better than another in terms of a certain measure, it may be actually worse in terms of the other measures. In this experiment, we used the first approach where we did not emphasize a certain evaluation measure but a trade-off among different measures, in which the best solution is selected among Pareto front solutions. The solutions in Pareto- front are ranked based on obtained evaluation measures, in other words the solutions are ranked in terms of each evaluation measure then the accumulative ranking is obtained by measuring average ranking. The solution with minimum average ranking is selected. The efficiency of LT is examined in the next sections.

## *10. Local Tuning vs. Global Tuning*

This section focuses on answering RQ2 by making comparison between LT and GT. The main difference between them is that LT attempts to find Pareto front solutions that improve accuracy for each project individually, whereas the GT attempts to find Pareto front solutions that improve accuracy for the whole dataset. In LT each project has different solutions than other projects in the same dataset, in other words, each project has different *k* value, set of features and weigh matrix. In GT, all projects share the same set of good solutions. So, the main theme of the comparison presented in this section is twofold: 1) to analyze whether the LT and GT methods can significantly improve the accuracy over ABE0. 2) To show which technique is more appropriate for the problem of adaptation. The ABE0 used in this comparison is the baseline ABE with best mean of nearest *k* analogies that minimize *MAE*. For each dataset, we run ABE0 repeatedly with changing *k* value from 1 to *n-1* every time. Then we select the ABE0 with *k* value that minimizes *MAE*. It is important to note that the best value of *k* is minimized globally (i.e. the *k* is same for all projects in the dataset). Finally, to make comparison between three variants we performed an empirical investigation over all 15 datasets using the experiment setup mentioned in Section 8.

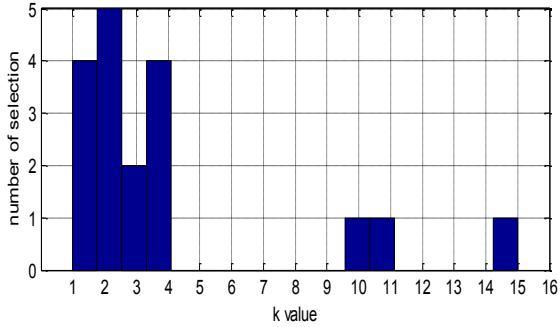
(a) Telecom dataset

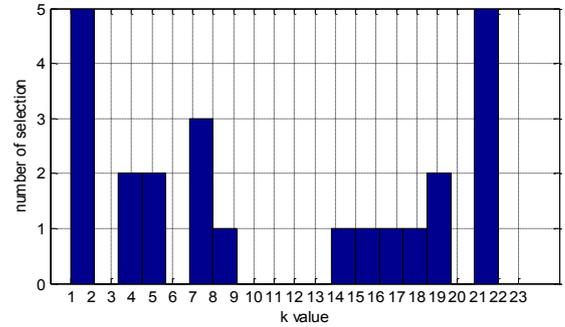
(b) Albrecht dataset

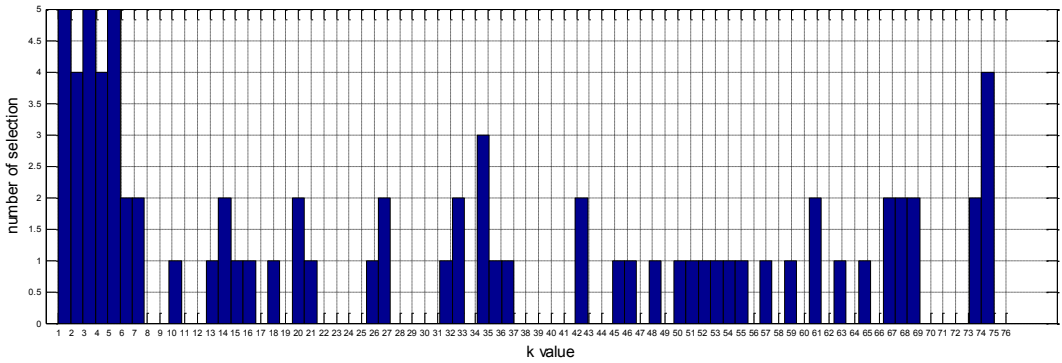
(c) desharnais dataset

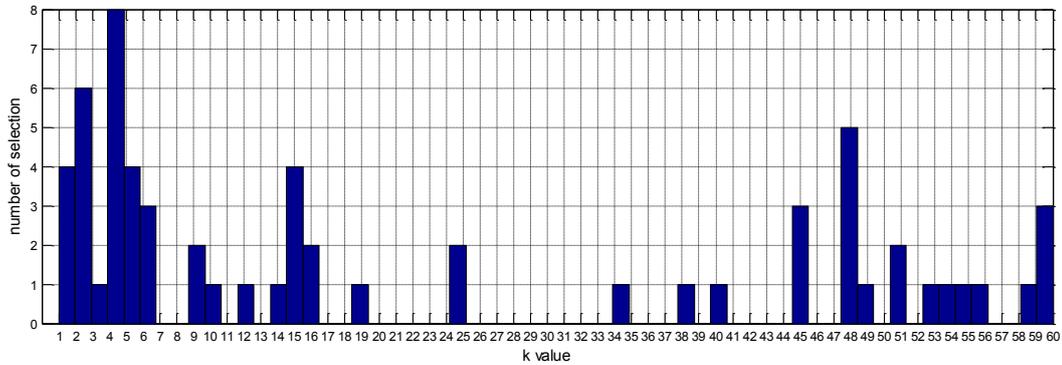
(d) maxwell dataset

Fig. 9 Bar chart of the best *k* for some datasets

Table 4 presents the results of the comparison between LT, GT and ABE0 in terms of *SA* and *Δ*. Although ABE0 was obtained based on minimizing *MAE* it cannot beat both LT and GT in terms of *SA* which is an indication to the performance of the optimization for adaptation procedure. Specifically, LT and GT achieve higher *SA* than ABE0, and both of them perform comparatively well, with superiority to LT as being optimized for each project individually. The effect size *Δ* in relation to random guess is frequently changed from small/medium to large, emphasizing the importance of using LT and GT to make adaptation. The *Δ* with value greater than 0.5 (medium effect size) is sufficient to conclude that the model does not generate its prediction by chance. The effect size test shows considerably large effect size over most datasets which confirms large effect improvement over guessing for both LT and GT. The cells in light grey in Table 4 represent the datasets for which some models could not generate estimates better than random guessing. It is worth noting that ABE0, LT and GT generate predictions by chances for around 50%, 25%, and 33% of the datasets respectively. In contrast to ABE0, the random predictions of LT and GT were frequently over the separated datasets such as desharnais_L1, desharnais_L2 and so forth. The primary reason for this worse behavior may be due to the heterogeneous structure of the separated datasets.

TABLE 4 *SA* and *Δ* between three variants

| Dataset | SA | | | Δ | | |
|---|---|---|---|---|---|---|
| | ABE0 | LT | GT | ABE0 | LT | GT |
| albrecht | 68.2 | 94.0 | 78.4 | 1.42 | 0.74 | 0.72 |
| kemerer | 50.6 | 69.6 | 50.6 | 0.5 | 0.68 | 0.6 |
| nasa | 69.8 | 94.8 | 85.1 | 1.9 | 2.3 | 2.1 |
| ISBSG | 35.1 | 67.6 | 62.8 | 0.48 | 0.71 | 0.66 |
| desharnais | 53.4 | 69.6 | 61.2 | 0.93 | 0.67 | 0.64 |
| desharnais_L1 | 32.8 | 55.5 | 40.9 | 0.48 | 0.58 | 0.45 |
| desharnais_L2 | 33.9 | 48.6 | 41.6 | 0.43 | 0.45 | 0.4 |
| desharnais_L3 | 39.0 | 45.4 | 38.2 | 0.63 | 0.52 | 0.5 |
| cocomo | 50.7 | 81.5 | 70.4 | 0.37 | 0.55 | 0.51 |
| cocomo_E | 28.2 | 46.1 | 43.0 | 0.29 | 0.4 | 0.48 |
| cocomo_O | 27.8 | 50.6 | 36.3 | 0.47 | 0.46 | 0.4 |
| cocomo_S | 18.8 | 50.2 | 46.8 | 0.18 | 0.39 | 0.35 |
| china | 69.4 | 75.7 | 76.3 | 0.94 | 0.53 | 0.5 |
| maxwell | 42.2 | 72.9 | 70.5 | 0.45 | 0.53 | 0.51 |
| telecom | 50.7 | 60.5 | 49.3 | 0.72 | 0.62 | 0.53 |

We also compare between three variants in terms of *MBRE*, *MIBRE* and *LSD*. These measures are considerably more reliable to compare between various predictions models than *MMRE* or *MdMRE* as explained in Section 4. In this section we show that the best Pareto solutions that consists of three decision variables can be used to improve performance of adaptation method in comparison to ABE0. Comparison with other adaptation methods is shown later in Section 13. Table 5 presents the predictive accuracy for each variant over all employed datasets. The cells in grey represent better accuracy but not necessarily statistically different. The overall results suggest that LT performs better than GT and ABE0 over all datasets, and GT still performs better than ABE0. Specifically, the results w.r.t *MBRE* shows that LT and GT generate more accurate predictions than ABE0 over all datasets, which indicates the performance of optimization and adaptation for improving accuracy of ABE. Although GT model finds the best solutions based on minimizing three evaluation measures: *MBRE*, *MIBRE* and *LSD* simultaneously, it rarely beats LT. Considering *MBRE* and *MIBRE* we can notice that LT wins 14 out 15 datasets whereas GT wins 1 out of 15. Considering *LSD*, we can notice that GT wins 3 out of 15 datasets whereas LT wins 12 out of 15 datasets. ABE0 has never outperformed other variant in any evaluation measure.

TABLE 5 Predictive accuracy for three variants

| Dataset | MBRE | | | MIBRE | | | LSD | | |
|---|---|---|---|---|---|---|---|---|---|
| | ABE0 | LT | GT | ABE0 | LT | GT | ABE0 | LT | GT |
| albrecht | 92.5 | **32.5** | 54.5 | 34.9 | **17.6** | 19.6 | 0.78 | 0.52 | **0.4** |
| kemerer | 75.9 | **17.2** | 60.3 | 32.1 | **12.1** | 21.3 | 1.2 | **0.63** | 0.67 |
| nasa | 112.6 | **7.2** | 21.7 | 34.4 | **6.1** | 16.4 | 0.8 | **0.11** | 0.24 |
| ISBSG | 93.4 | **33.3** | 35.2 | 38.7 | **15.9** | 18.0 | 0.7 | **0.39** | 0.4 |
| desharnais | 61.9 | **36.4** | 45.3 | 30.8 | **15.4** | 20.2 | 0.52 | **0.43** | 0.46 |
| desharnais_L1 | 107.9 | **55.6** | 70.5 | 38.2 | **20.4** | 24.9 | 0.77 | **0.53** | 0.62 |
| desharnais_L2 | 94.7 | **53.5** | 60.6 | 33.9 | **19.9** | 21.2 | 0.74 | **0.53** | 0.57 |
| desharnais_L3 | 77.2 | **53.1** | 56.0 | 30.0 | **23.4** | 25.6 | 0.66 | 0.54 | **0.52** |
| cocomo | 506.1 | **63.2** | 84.9 | 55.9 | **24.8** | 31.3 | 1.51 | **0.55** | 0.66 |
| cocomo_E | 1728.6 | 1581.2 | **1299.8** | 64.5 | **52.2** | 53.4 | 2.7 | 2.0 | **1.87** |
| cocomo_O | 1142.4 | **1070.0** | 3264.0 | 63.0 | **44.2** | 44.9 | 2.0 | **1.7** | 2.2 |
| cocomo_S | 3229.2 | **1710.0** | 2326.4 | 75.5 | **45.9** | 48.7 | 2.11 | **1.9** | 2.0 |
| china | 61.2 | **45.1** | 50.5 | 29.4 | **17.1** | 19.6 | 0.57 | **0.47** | 0.49 |
| maxwell | 165.2 | **30.1** | 35.7 | 42.5 | **17.6** | 20.9 | 0.89 | **0.34** | 0.37 |
| telecom | 79.6 | **33.2** | 73.9 | 35.4 | 31.1 | **16.6** | 0.67 | **0.41** | 0.64 |

It is important to note that the performance of three variants over separated datasets are very bad with large error deviations. Splitting large datasets such as cocomo and desharnais into coherent small datasets did not improve predictive accuracy. This suggests that separating dataset has greater impact on the structure of that datasets and may lead to worse performance as in our case. In summary, we can notice that the procedure of finding individual best solution for each project would generally perform efficiently than finding shared best solution for all project in the datasets.

Based on the above comparison, we can conclude that both LT and GT perform better than ABE0, and particularly LT performs better than GT. To make this hypothesis more well-grounded, we did statistical significance tests over all datasets. The Wilcoxon rank-sum statistical tests, in Table 6, at the overall level of significance of 0.05 for comparing all variants detect statistically significant difference for most datasets. The comparison between the predictions generated by LT and ABE0 suggest that there are significance differences between them over all datasets except some separated datasets such as desharnais_L3 and cocomo_O. Likewise, the comparison between ABE0 and GT suggest that there is no significance difference between them over some separated datasets. On the other hand, we cannot find any significance difference between predictions generated by LT and GT over small datasets such as albrecht, kemerer, nasa and telecom, and over separated datasets.

## 11. The importance of feature optimization

As shown in Section 10, LT and GT can be used to automatically find the best decision variables that provide a good trade-off among different evaluation measures, so that the manager does not need to decide on how many analogies should be used or which feature set are more accurate results. This section concentrates on answering RQ3 which states that whether the use of all features with LT and GT have room for further improvements. It is already recognized that using subset of features would perform better than using all features in terms of evaluation measures [8]. But this would arise another concern about how these features should be found. Frequently, previous studies showed that the best feature set that is used to improve ABE0 is frequently found based on minimizing one of the evaluation measures and subject to the decision made by an expert such as similarity measure and data standardization. So these features would perform better in terms of that evaluation measure but worse in terms of others. In previous section, we showed that optimizing features with other decision variables simultaneously based on many evaluation measures work well, especially if they are evaluated by other evaluation measures.

TABLE 6 Statistical significance test between three variants

| Dataset | p-value | | |
|---|---|---|---|
| | ABE0 Vs. LT | ABE0 vs. GT | LT vs. GT |
| albrecht | **0.0013** | **0.006** | 0.74 |
| kemerer | **0.045** | **0.04** | 0.95 |
| nasa | **0.0001** | **0.003** | 0.026 |
| ISBSG | **0** | **0** | **0.01** |
| desharnais | **0** | **0.001** | **0.005** |
| desharnais_L1 | **0** | **0.003** | 0.47 |
| desharnais_L2 | **0.009** | **0.028** | 0.56 |
| desharnais_L3 | 0.57 | 0.79 | 0.82 |
| cocomo | **0** | **0.0002** | **0.032** |
| cocomo_E | **0.2** | 0.37 | 0.6 |
| cocomo_O | 0.14 | 0.19 | 0.97 |
| cocomo_S | 0 | 0.001 | 0.58 |
| china | 0 | 0 | 0.046 |
| maxwell | 0 | 0 | 0.01 |
| telecom | 0.008 | 0.011 | 0.78 |

In this section we need to investigate that does the features optimization really contribute towards this good performance? To answer this question we re-run the LT and GT but with keeping the feature decision variable $v$ unchanged during optimization process (i.e. $v=2^M-1$) whereas the other decision variables (i.e. $k$ and $w$) change normally during the run. This allows to see the contribution of feature optimization on the final accuracy. So if the results when using all features do not significantly improve the prediction accuracy, we can definitely confirm that optimizing features within LT or GT is considerably more accurate. The new adaptation strategy without feature optimization will be called hereafter LT* and GT*.

Table 7 shows the prediction results in terms of four evaluation measures in addition to the effect size for both LT* and GT*. The effect size of LT* and GT* is computed with relation to LT and GT respectively. For example, the Eq. 21 would appear as shown in Eq. 23 for LT*. This allows us to justify if there is large effect improvement over LT and GT since the statistical significance test alone is not so informative if both predictions models are significantly different.

$$\Delta_{LT^*} = \frac{MAE_{LT^*} - \overline{MAE}_{LT}}{SP_{LT}} \qquad (23)$$

The obtained results in terms of *SA* shows that both LT* and GT* are predicting over all datasets since they fall comfortably beyond the *MAE* of random guessing. So we can believe that the prediction generated by LT* or GT* are highly unlikely to have arisen by chance. However, when comparing LT* and GT* to LT and GT w.r.t *SA* we can find LT and GT are more accurate which suggests that using the optimized features is better than using all features. To confirm that we computed effect size for both LT* and GT* with relation to LT and GT respectively. We can notice that none of LT* and GT* has a medium effect size over any dataset. This alone should suggest we cannot find significant improvements when using all features either with GT* or LT*.

On the other hand, the accuracy values in terms of *MBRE*, *MIBRE* and *LSD* are also going into that direction. If we look closer at the obtained values and compare it to the results obtained in Table 5 we can find that LT and GT performs better with optimizing feature set. This finding emphasize more on the importance of feature optimization together with other decision variables on the final accuracy. Another important result is that using all feature would also perform better than ABE0.

TABLE 7 Predictive accuracy for LT* and GT*

| Dataset | SA | | Δ | | MBRE | | MIBRE | | LSD | |
|---|---|---|---|---|---|---|---|---|---|---|
| | LT* | GT* | LT* | GT* | LT* | GT* | LT* | GT* | LT* | GT* |
| albrecht | 77.5 | 52.6 | 0.69 | 0.76 | 66.4 | 97.3 | 27.5 | 39.7 | 0.55 | 0.89 |
| kemerer | 56.0 | 36.1 | 0.4 | 0.3 | 52.8 | 75.4 | 20.6 | 31.1 | 0.76 | 0.69 |
| nasa | 90.4 | 61.9 | 2.2 | 1.83 | 10.83 | 86.6 | 9.91 | 61.9 | 0.16 | 0.83 |
| ISBSG | 42.3 | 32.1 | 0.38 | 0.33 | 154.3 | 119.6 | 24.5 | 36.11 | 0.73 | 0.96 |
| desharnais | 56.7 | 55.2 | 0.59 | 0.99 | 50.1 | 54.1 | 19.9 | 28.8 | 0.49 | 0.52 |
| desharnais_L1 | 41.2 | 12.3 | 0.44 | 0.14 | 79.9 | 153.2 | 26.1 | 45.5 | 0.63 | 1.08 |
| desharnais_L2 | 42.7 | 30.2 | 0.39 | 0.39 | 66.9 | 96.6 | 20.7 | 35.2 | 0.58 | 0.72 |
| desharnais_L3 | 44.2 | 29.9 | 0.57 | 0.47 | 59.2 | 88.3 | 26.9 | 35.7 | 0.55 | 1.28 |
| cocomo | 82.4 | 48.2 | 0.47 | 0.32 | 70.2 | 395.7 | 28.8 | 57.0 | 0.6 | 1.8 |
| cocomo_E | 42.6 | 29.8 | 0.43 | 0.3 | 1659.5 | 2705 | 56.8 | 65.9 | 2.2 | 3.24 |
| cocomo_O | 45.4 | 26.4 | 0.73 | 0.4 | 1610 | 3317 | 54.4 | 56.0 | 1.77 | 2.87 |
| cocomo_S | 49.6 | 44.1 | 0.36 | 0.35 | 1754 | 4869.8 | 48.8 | 77.3 | 2.0 | 2.85 |
| china | 72.2 | 72.9 | 0.64 | 0.59 | 147.9 | 62.4 | 23.8 | 21.9 | 0.74 | 0.53 |
| maxwell | 64.4 | 52.5 | 0.48 | 75.6 | 41.9 | 97.3 | 22.8 | 39.7 | 0.42 | 0.87 |
| telecom | 58.6 | 28.9 | 0.58 | 0.54 | 42.3 | 84.9 | 29.4 | 27.3 | 0.46 | 0.86 |

The results without statistical significance test would be not convincing so we make comparison between each two variants (i.e. LT vs LT* and GT vs GT*). The results of statistical significance test based on absolute errors are shown in Table 8. Surprisingly, we cannot find any significance difference between LT against LT* or GT against GT* over the separated datasets. On the other hand we can see that there is significance difference between LT and its counterpart LT*, and between GT and its corresponding counterpart GT* over most datasets. This indicates that both of them generate different predictions with superiority to LT and GT since they yield better performance w.r.t to three evaluation measures. So we believe that there is sufficient evidence that using different feature set for each project is more efficient than using all features as confirmed by the comparison between LT and LT*. Moreover, optimizing feature set for the whole dataset is also better than using all features as shown in the comparison between GT and GT*.

TABLE 8 Statistical significance test between counterparts

| Dataset | p-value | |
|---|---|---|
| | LT* Vs. LT | GT* vs. GT |
| albrecht | **0.032** | **0.001** |
| kemerer | **0.048** | **0.08** |
| nasa | **0.03** | **0.03** |
| ISBSG | **0.001** | **0.001** |
| desharnais | **0.02** | **0.004** |
| desharnais_L1 | 0.08 | 0.06 |
| desharnais_L2 | 0.09 | 0.05 |
| desharnais_L3 | 0.73 | 0.07 |
| cocomo | **0.04** | **0.003** |
| cocomo_E | 0.22 | 0.4 |
| cocomo_O | 0.52 | 0.85 |
| cocomo_S | 0.14 | **0.01** |
| china | **0.01** | **0.001** |
| maxwell | **0.02** | **0.001** |
| telecom | **0.03** | **0.04** |

## *12. The importance of weighting optimization in the adaptation strategy*

This section concentrates on answering question RQ4 which states that: does the use of weighting values contribute towards improving prediction accuracy of the adaptation technique? It is already known that using weighting mechanism shows considerable performance when it is applied in project retrieval and some adaptation methods such as AQUA and GA [1]. Usually software managers tend to use simple weighting mechanism that is feasible and easy to apply such as inverse ranked weighted mean or similarity between projects. Although these mechanisms follow formal procedures in finding weight values, they are considered useful only when the dataset structure is rather simple and normally distributed. The weight values that found by LT or GT are generated randomly and do not follow a particular algorithm, then are changed according to the best positions of particles. To better understand the importance of weighting in our adaptation strategy we aim at comparing the use of non-weighted form of our adaptation strategy with the weighted version (i.e. original LT or GT).

To do so, we re-run the LT and GT but without considering weight values (i.e. using equal weights). The modified variant of both LT and GT will be called hereafter LT+ and GT+ consequentially. Similar to previous section, weighting values remain equal and unchanged (i.e. $w_{ij} = 1, \forall i = 1,2,...,n; \forall j = 1,2,...,M$) during the optimization process while other decision variables ($k$ and $v$) change normally according to the optimization algorithm. So, the main objective of the comparison presented in this section is to analyze the level of accuracy improvements when optimizing adaptation weights.

First, let's analyze the accuracy values considering all data sets. Table 9 shows the predictive performance and effect size for the Pareto efficient adaptation method without considering the weight matrix. The results of *SA* shows that LT+ and GT+ do not frequently generate successful predictions than random guessing which suggest that they are not reliable to ignore the role of weighting mechanism. Moreover, the SA of LT+ and GT+ are worse than those of LT and GT over all datasets. Also, we are interested with the results of effect size in comparison to the original LT and GT. So we measure the effect size for LT+ and GT+ with relation to LT and GT as baseline respectively. The effect size results were almost small (i.e. $\Delta \approx 0.2$) suggesting that small effect size improvements than LT or GT. Likewise, the performance figures in terms of *MBRE* and *MIBRE* and *LSD* suggest that ignoring the weight values did not contribute significantly towards accuracy improvements of LT and GT. The value of these evaluation measures become worse when we use LT+ and GT+ with poor *MBRE* and *MIBRE* and relatively bad *LSD* in comparison to LT and GT respectively. So we can believe that the optimization of three decision variables simultaneously based on optimizing various evaluation measures allow software manager to save time in finding the appropriate decisions in very large space of configurations.

TABLE 9 Predictive accuracy for LT+ and GT+

| Dataset | SA | | Δ | | MBRE | | MIBRE | | LSD | |
|---|---|---|---|---|---|---|---|---|---|---|
| | **LT+** | **GT+** | **LT+** | **GT+** | **LT+** | **GT+** | **LT+** | **GT+** | **LT+** | **GT+** |
| albrecht | 82.2 | 76.2 | 0.86 | 0.7 | 67.3 | 83.9 | 22.6 | 30.2 | 0.57 | 0.68 |
| kemerer | 56.3 | 50.7 | 0.41 | 0.64 | 51.55 | 83.9 | 16.4 | 36.1 | 0.66 | 0.71 |
| nasa | 82.3 | 79.0 | 0.93 | 0.83 | 27.4 | 50.4 | 20.3 | 24.0 | 0.28 | 0.5 |
| ISBSG | 60.7 | 23.7 | 0.63 | 0.23 | 37.8 | 119.1 | 19.2 | 35.6 | 0.41 | 0.88 |
| desharnais | 65.6 | 43.3 | 0.68 | 0.63 | 39.8 | 63.4 | 18.4 | 30.9 | 0.48 | 0.55 |
| desharnais_L1 | 38.4 | 33.4 | 0.4 | 0.22 | 70.7 | 224.7 | 25.2 | 49.6 | 0.62 | 1.4 |
| desharnais_L2 | 39.1 | 28.55 | 0.35 | 0.35 | 60.6 | 154.0 | 21.5 | 36.1 | 0.57 | 1 |
| desharnais_L3 | 39.6 | 31.6 | 0.51 | 0.33 | 55.9 | 205.4 | 27.2 | 50.5 | 0.59 | 1.31 |
| cocomo | 80.9 | 52.0 | 0.46 | 0.4 | 66.1 | 319.4 | 29.0 | 55.5 | 0.62 | 1.6 |
| cocomo_E | 44.9 | 13.7 | 0.39 | 0.1 | 1304.1 | 4273.0 | 53.3 | 80.6 | 2.3 | 5.9 |
| cocomo_O | 50.4 | 11.4 | 0.78 | 0.11 | 1070.1 | 1103.1 | 44.9 | 65.6 | 1.9 | 2.88 |
| cocomo_S | 50.5 | 8.4 | 0.39 | 0.04 | 1711.1 | 1776.4 | 46.2 | 53.7 | 1.91 | 2.8 |
| china | 75.2 | 79.7 | 0.52 | 1.1 | 45.9 | 55.2 | 17.3 | 24.1 | 0.48 | 0.51 |
| maxwell | 72.4 | 41.6 | 0.54 | 0.41 | 32.2 | 105.8 | 24.2 | 40.2 | 0.41 | 0.8 |
| telecom | 11.3 | 42.6 | 0.11 | 0.5 | 82.1 | 83.5 | 32.1 | 35.0 | 0.71 | 0.72 |

Even though the effect size suggest bad performance of Non-weighted LT+ and GT+ in comparison to LT and GT, we are still interested to see the p-value of statistical significance test over all datasets. Table 10 shows the p-values of the comparison between each two corresponding opponents LT vs. LT+ and GT vs. GT+ using Wilcoxon sum rank test and based on absolute prediction errors. The general trend shows that there is a significant difference between predictions generated by LT and LT+ and between GT and GT+, which indicates that both LT and GT are able to produce better accuracy. So we believe that the weight values is more important for our adaptation strategy. Nevertheless, even though the LT and GT produce meaningful predictions over LT+ and GT+, we could not find any difference between them over some separated datasets. So we can see that the use of separated datasets is still problematic and did not contribute well in improving accuracy than original dataset.

TABLE 10 Statistical significance test between each two opponent

| Dataset | p-value | |
|---|---|---|
| | LT+ Vs. LT | GT+ vs. GT |
| albrecht | **0.01** | **0.01** |
| kemerer | **0.01** | **0.02** |
| nasa | **0.01** | **0.036** |
| ISBSG | **0.031** | **0.002** |
| desharnais | **0.04** | **0.001** |
| desharnais_L1 | **0.04** | 0.12 |
| desharnais_L2 | 0.09 | **0.01** |
| desharnais_L3 | 0.11 | **0.037** |
| cocomo | **0.001** | **0.02** |
| cocomo_E | 0.32 | 0.27 |
| cocomo_O | 0.39 | 0.47 |
| cocomo_S | 0.28 | 0.35 |
| china | **0.01** | **0.033** |
| maxwell | **0.001** | **0.001** |
| telecom | **0.001** | **0.04** |

## 13. Further Analysis

This section presents performance figures of LT and GT against various adaptation techniques used with ABE. The adaptation methods that have been compared are: LSE, RTM, GA and AQUA as explained in Section 2. The objective of this comparison is to ensure that the proposed adaptation strategy works comparatively well against other adaptation methods existing in literature. To make this comparison much well grounded, we used the same validation procedure followed by LT and GT (i.e. Leave-one-out cross validation). Since the selection of the best *k* setting in other adaptation methods is not dynamic, there was necessarily to find the best *k* value that almost fits each model, therefore we perform empirical validation for each method to find the best *k* value that yield minimum *MAE*. The principal reason for this process is to ensure all methods undergo the same assumption since the original experiments of those studies found the best *k* by minimizing *MMRE*, which is not reliable anymore to be relied on. Thus, the best variant of each method with best *k* value has been used in the further empirical comparison.

Table 11 summarize the resulting performance figures for all investigated adaptation methods. The most successful method should have lower *MBRE*, *MIBRE*, *LSD*. The grey cells with superscripts * represent the model and dataset for which GT only has been outperformed. The grey cells with superscripts ** represent the model and dataset for which LT and GT have been outperformed together. If we look closer at the results we can find that LT and GT are rarely beaten by any of the adaptation methods except over some datasets such as china and nasa.

However, these findings are indicative of the superiority of LT and GT in optimizing the adaption decision variables, and consequentially improve overall predictive performance of ABE. One of the advantages of using LT and GT is that they are fully automated procedure because they do not need human intervention to guess how many nearest analogies should be used or which features should be involved. Also from the obtained results we can observe that there is evidence that our adaptation method can work better for datasets with discontinuities (e.g. Maxwell, and COCOMO). We speculate that prior Software Engineering researchers who failed to find best *k* setting, did not attempt to optimize three decision variables simultaneously with the adaptation method itself for each individual project before building the model.

TABLE 11 Predictive accuracy for various adaptation methods

| Dataset | MBRE | | | | MIBRE | | | | LSD | | | |
|---|---|---|---|---|---|---|---|---|---|---|---|---|
| | LSE | RTM | GA | AQUA | LSE | RTM | GA | AQUA | LSE | RTM | GA | AQUA |
| albrecht | 85.7 | 85.1 | 166.4 | 90.1 | 34.1 | 30.9 | 44.4 | 38.0 | 0.7 | 0.7 | 1.1 | 0.7 |
| kemerer | 71.4 | 66.0 | 74.7 | 75.9 | 29.8 | 33.9 | 32.4 | 32.1 | 0.7 | 0.6** | 0.7 | 0.7 |
| nasa | 27.1 | 20.8* | 102.7 | 109.2 | 20.1 | 15.5* | 34.5 | 35.2 | 0.3 | 0.2* | 0.8 | 0.8 |
| ISBSG | 119.1 | 140.7 | 113.8 | 122.6 | 35.6 | 42.9 | 35.0 | 38.4 | 0.9 | 0.8 | 0.9 | 0.94 |
| desharnais | 59.8 | 63.5 | 83.4 | 83.6 | 31.5 | 31.0 | 36.7 | 36.7 | 0.5 | 0.5 | 0.7 | 0.7 |
| desharnais_L1 | 247.2 | 152.0 | 202.7 | 202.7 | 48.9 | 44.1 | 54.7 | 54.7 | 1.5 | 0.9 | 1.3 | 1.3 |
| desharnais_L2 | 111.6 | 104.8 | 165.4 | 165.7 | 33.9 | 36.1 | 41.2 | 41.2 | 0.9 | 0.8 | 1.1 | 1.1 |
| desharnais_L3 | 256.0 | 113.0 | 171.2 | 171.0 | 59.0 | 37.7 | 44.3 | 44.2 | 1.6 | 0.8 | 1.2 | 1.2 |
| cocomo | 111.0 | 119.4 | 465.6 | 473.5 | 40.5 | 43.6 | 57.8 | 57.8 | 0.9 | 0.8 | 1.8 | 1.9 |
| cocomo_E | 16861.6 | 2727.0 | 3919.9 | 3913.1 | 81.1 | 72.0 | 80.1 | 80.1 | 7.6 | 3.0 | 5.8 | 5.8 |
| cocomo_O | 2056.3* | 22565.0* | 1173.7* | 1175.4* | 73.8 | 67.3 | 69.6 | 69.6 | 2.5 | 3.0 | 3.2 | 3.2 |
| cocomo_S | 8867.0 | 7596.0 | 1546** | 15434** | 67.3 | 74.7 | 55.7 | 56.1 | 4.3 | 3.4 | 2.8 | 2.8 |
| china | 25.1** | 25.0** | 68.5 | 68.7 | 14.1** | 14.1** | 31.5 | 31.6 | 0.3** | 0.3** | 0.6 | 0.6 |
| maxwell | 105.8 | 79.8 | 271.6 | 271.7 | 40.2 | 38.3 | 57.5 | 57.5 | 0.8 | 0.6 | 1.5 | 1.5 |
| telecom | 92.0 | 85.8 | 74.6 | 74.8 | 38.8 | 34.0 | 32.0 | 32.1 | 0.8 | 0.7 | 0.7 | 0.7 |

To summarize the results we consult *win, tie, loss* algorithm [25] to show the complete picture of our analysis to compare the predictive performance of the variants of Adaptation methods as suggested by Kocaguneli et al. [25]. As a consequence, all six adaptation methods have been first evaluated over 15 datasets using Leave-one-out cross validation and four evaluation measures. For each variant of adaptation method, we record *MAE*, *MBRE*, *MIBRE* and *LSD*. We also consult *win, tie, loss* algorithm to compute *win-loss* (i.e. *win* minus *loss*) for each variant after comparing all variants with each other across all error measures. All single methods are ranked with respect to all error measures in addition to (*win-loss*) over all datasets which resulted in (15 × 4 × 5) = 300 possible rankings. Figure 10 shows the adaptation methods variants, sorted by the calculated number of losses seen in all evaluation measures and all datasets. The adaptation method with the largest score is ranked #1 which is LT. At the other end of the scale, the adaptation method with the lowest score is ranked #6 which is AQUA. Notice that, all methods are ranked in ascending order (i.e. lowest first) over all error measure. It is clear from final scores that LT and GT variants are the top winners and they also ranked top across all other error measures. The method LT is ranked first among all methods based on accumulative decision, but this method has been the winner over only three error measures (*LSD*, *MBRE* and *MIBRE*). Apparently, there is significant difference between the best and worst methods in terms of number of losses (in the extreme case it is close to 125). The *win-tie-loss* results offer yet more evidence for the superiority of LT over other adaptation techniques. Also the obtained *win-tie-loss* results confirmed that the predictions based on LT and GT methods presented statistically significant but necessarily accurate estimations than other methods. This suggests that there is a significant difference if the prediction generated by LT or GT against other models.

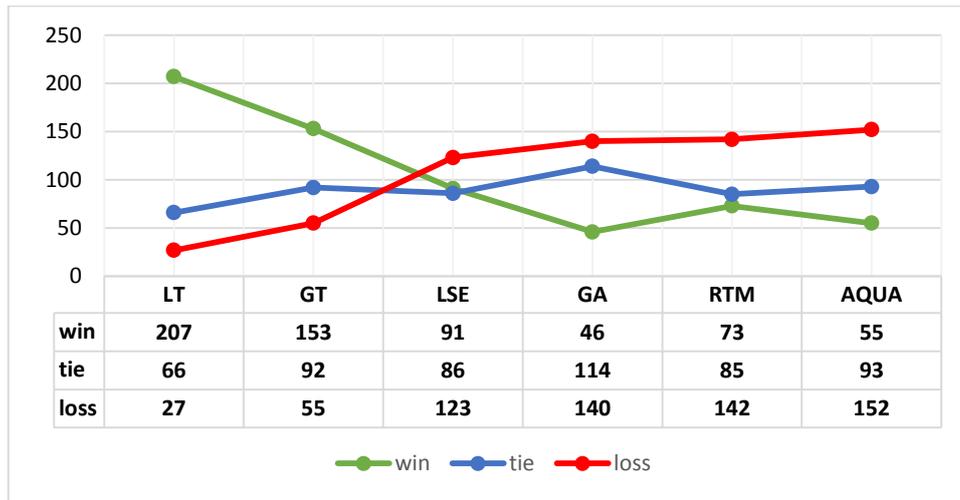

Fig. 10 *win-tie-loss* results for all models.

Although the GA method is one of the most efficient optimization techniques, it occupied the late positions with the poor performance across multiple evaluation measures. On contrast, the comparative performance of LSE method can be attributed to the fact that it uses the size feature as the adaptation factors, in which this feature is considered the strongest correlated feature with effort in all datasets.

TABLE 12 Ranking summary for different adaptation methods over three evaluation measures (*MBRE*, *MIBRE* and *LSD*)

|  | MBRE | | MIBRE | | LSD | |
| --- | --- | --- | --- | --- | --- | --- |
|  | SD | Average | SD | Average | SD | Average |
| LT | 0.7 | 1.3 | 1.8 | 3.4 | 1.6 | 2.2 |
| GT | 1.1 | 2.5 | 1.5 | 3.4 | 1.8 | 2.3 |
| LSE | 1.3 | 4.5 | 1.9 | 3.4 | 1.7 | 4 |
| GA | 1.2 | 4.4 | 1.8 | 3 | 1.5 | 2.8 |
| RTM | 1.4 | 3.6 | 1.7 | 3.8 | 1.9 | 3.2 |
| AQUA | 1.5 | 4.7 | 1.6 | 4 | 1.7 | 4.2 |

Recall Kocaguneli et al. [25] results, any method is to be superior to others, should be ranked first and has a minimum number of changes in their ranks. Therefore we used Standard Deviation (SD) of rank changes to measure stability of ranking for each method across different experimental conditions. The minimum SD represents more stable ranking because it has smaller ranking distribution.

Table 12 shows the success of adaptation methods through ranking over *MBRE*, *MIBRE*, and *LSD* consequentially. It shows the SD and average of ranking across multiple datasets. The most successful models have higher ranking and lower rank change which is indication to the stability of ranking. The 'method rank per dataset' has been obtained by ranking every method based on different error measure (i.e. ranking based *MBRE*, ranking based *MIBRE* and *LSD*) then we aggregate those ranks using across all datasets. Looking carefully at this Table we can notice that the winner over all datasets in terms of *MBRE* was LT with stable ranking (i.e. rank#1 over all datasets). Other adaptation have instability in ranking which conclude that it was previously hard to identify the superior variant among those historical adaptation methods. While some methods have large amount of changes, other do not. Better yet, as shown in the Figure 10, the LT had the lowest rank change seen in any of other 5 adaptation techniques. Although the RTM has stable ranking with smaller amount of rank changing (i.e. zero) but it ranks lower than other models with bad performance according to all error measures.

Figure 11 sorts all 6 methods according to their SD of rank changes over three evaluation measures mentioned in Table 12. The y-axis shows the SD of rank changes obtained for each adaptation method, as we compare the ranks across all evaluation measures and over multiple datasets. For example, the top ranked method of LT has nearly 1.1 SD in difference between best and worst ranks. A line drawn parallel to x-axis at y = 1.3 gives methods, whose SD of rank change is less/more than 1.3 whereas the line drawn parallel to y-axis exactly after GT gives methods whose ranked top and have lowest stable rank change. From this figure, we can observe that the LT and GT methods in table 12 have lowest rank changes than others, which is good news since the lowest rank change is an indication of the ranking stability for that method. Notably, RTM is relatively considered stable as they have narrower SD of rank changes but unfortunately they are far from the top ranked methods. Thus, even that these top-ranked methods jumped rank by their maximum change, they would still be working better than most of the others 3 methods. However, LSE, GA and AQUA ranked between #3 and #6 (inclusive) have SD of rank changes above 1.6, i.e. they are "unstable" in this region.

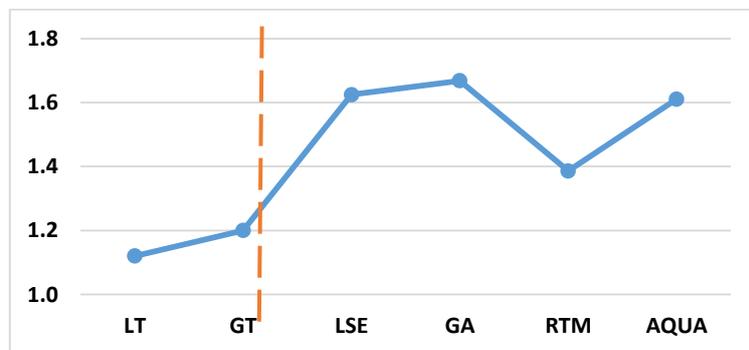

Fig. 11 Ranking changes across multiple datasets

Other results offer yet more evidence of the performance of LT and GT over existing adaptation methods. Figure 12 depicts plot of the Bonferroni test for ranking methods over three evaluation measures (*BRE*, *IBRE* and *AE*). We use the Bonferroni-Dunn test to compare the ranking of LT and GT methods against various adaptation methods. The plots have been obtained after applying ANOVA test followed by Bonferroni test. The ANOVA test results in p-value close to zero (i.e. <0.01) which implies that the difference between rankings of LT and GT against other adaptation methods are statistically significant across the three utilized error measures. The horizontal axis in these figures corresponds to the average rank of across different datasets. The more this line is situated to the left, the better performing the method is. The dotted vertical lines in the figures indicate the critical difference at the 95% confidence level. Obviously, the LT and GT occupy better ranks than all adaptation methods including GT in terms of all measures with preference to LT as significantly better than GT in terms of *BRE* and *IBRE*. Nevertheless, there was no significance difference between LT and GT in terms of *AE*.

## 14. Conclusions and Future Work

Finding the appropriate decisions to adapt the nearest analogies in ABE method is relatively a nontrivial task. Prior methods have attempted to use these decisions or part of them manually based on expert opinion. This approach has some limitations because the expert cannot identify all possible decision combinations that may reach thousands. In this paper, we proposed two adaptation methods (LT and GT) based on using multi-objective optimization algorithm MOPSO to identify optimal decision variables ($k$, $v$, and $w$) that present a good trade-off between different evaluation measures. In our study, we defined four research question to study the performance of the proposed adaption methods and investigate the impact of decision variables on the optimization process.

Based on our analysis, we summarize the following findings:

1. The use of LT tends to be more efficient than using GT in most datasets. In other words, optimizing all decision variables for each project individually produce better results than optimizing them for the whole dataset. We have shown in section 10 that LT statistically generates better predictions than those of GT over most datasets.
2. Optimizing all three decision variables of adaptation method simultaneously is better than optimizing part of them. We have shown in sections 11 and 12 that fixing any one of the decision variable does not improve predictive performance and makes the adaption method to generate worse estimates. This finding has been statistically tested using Wilcoxon sum rank test.
3. Remarkably, the LT and GT work better than other existing adaptation methods as confirmed in Figures 10 and 12.
4. Most adaption methods did not yield good results over the separated datasets comparing to the results of their original datasets. This emphasizes us to recommend not separating dataset based on categorical features. Even though, the LT and GT still perform better than other adaptation methods over these datasets.

Apart from trying to address the reported issues regarding previous ABE studies in terms of dataset and model, we also tried to meet the methodological problems such as testing only on a limited number of datasets and lacking statistical checks on the results. Therefore, we utilized various datasets from multiple resources and evaluated our results on the basis of Wilcoxon signed rank test at a 95% confidence level. Furthermore, rather than proposing a best solution a priori as the traditional ABE methods do, what LT and GT do is automatically identify some Pareto front solutions that make trade-off among different evaluation measures. So we don't emphasize any particular evaluation measures as in previous studies. Therefore all results obtained support this attitude and show better accuracy than other competitors over all evaluation measures.

A future work is planned to study the use of MOPSO to select the best design decision of ABE. As we have pointed out in the introduction section the ABE has many design decisions and project manager has no time to find which design decision that best fits his/her data. So the next study aims to evaluate the performance of MPOSO on finding the right decision for the data in hand using multiple evaluation measures.

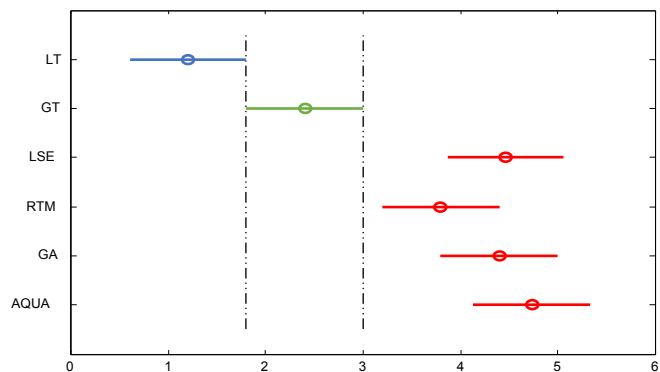

(a) *BRE*

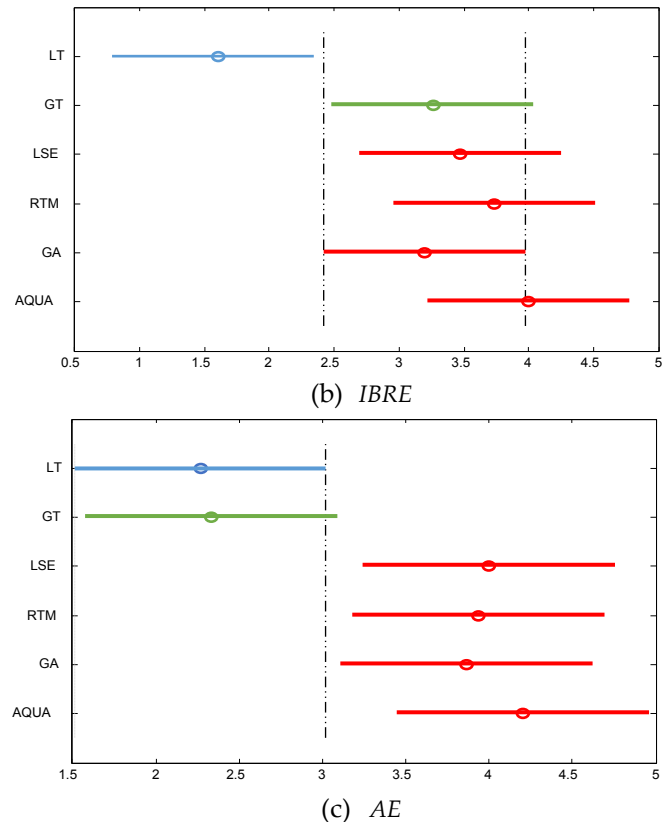

Fig. 12 Ranking of variants of adaptation methods for three evaluation measures: (a) *BRE*, (b) *IBRE*, and (c) *AE*. The dashed line the 95 percent significance level. (a) Plot of the Bonferroni-Dunn test for *BRE*, (b) plot of the Bonferroni-Dunn test for *IBRE*, (c) plot of the Bonferroni-Dunn test for *AE*

**15. Acknowledgements**

The authors are grateful to the Applied Science Private University, Amman, Jordan, for the financial support granted to this research project.